\PassOptionsToPackage{table,dvipsnames}{xcolor}
\documentclass[acmtog,anonymous=false,review=false]{acmart}

\usepackage{booktabs} 

\setcitestyle{square}

\acmSubmissionID{868}

\usepackage{amsbsy}
\usepackage[mathscr]{euscript}
\usepackage{amsmath}
\usepackage{stackrel}
\usepackage{nicefrac}

\usepackage{pifont}
\newcommand{\cmark}{\ding{51}}%
\newcommand{\xmark}{\ding{55}}%

\usepackage{booktabs}
\usepackage{colortbl}
\usepackage{array}
\usepackage{ragged2e}
\usepackage{tabularx}

\usepackage{makecell}
\usepackage{multirow}

\usepackage{graphicx}
\usepackage{siunitx}
\newcommand{\greencheck}{{\color{Green}\cmark}\xspace}
\newcommand{\green}{\cellcolor{Green!12.5}\greencheck}
\newcommand{\yellowcheck}{{\color{YellowOrange}(\cmark)}\xspace}
\newcommand{\yellow}{\cellcolor{YellowOrange!12.5}\yellowcheck}
\newcommand{\redcheck}{{\color{red}\xmark}\xspace}
\newcommand{\red}{\cellcolor{red!12.5}\redcheck}

\usepackage{algpseudocode}
\usepackage[ruled]{algorithm2e} 

\SetAlFnt{\small}
\SetAlCapFnt{\small}
\SetAlCapNameFnt{\small}
\SetAlCapHSkip{0pt}
\IncMargin{-\parindent}

\usepackage{dblfloatfix}
\usepackage{float}

\usepackage{array}

\newcolumntype{L}[1]{>{\raggedright\let\newline\\\arraybackslash\hspace{0pt}}m{#1}}
\newcolumntype{C}[1]{>{\centering\let\newline\\\arraybackslash\hspace{0pt}}m{#1}}

\usepackage{siunitx}
\usepackage{subcaption}

\acmJournal{TOG}
\acmVolume{0}
\acmNumber{0}
\acmArticle{0}
\acmYear{2023}
\acmMonth{0}

\newcommand\rev[1]{\textcolor{black}{#1}}

\newcommand{\argmin}[1]{\operatorname*{arg\,min}_{\{ #1 \}}}

\begin{document}

\title{Thin On-Sensor Nanophotonic Array Cameras}

\author{Praneeth Chakravarthula}
\authornote{Indicates equal contribution.}
\email{cpk@cs.unc.edu}

\author{Jipeng Sun}
\authornotemark[1]
\email{jipeng.sun@princeton.edu}
\affiliation{%
  \institution{Princeton University}
  \streetaddress{35 Olden St}
  \city{Princeton}
  \state{New Jersey}
  \country{USA}
  \postcode{08540}
}
\author{Xiao Li}
\email{xl2710@princeton.edu}
\author{Chenyang Lei}
\email{leichenyang7@gmail.com}
\author{Gene Chou}
\email{gchou@princeton.edu}
\author{Mario Bijelic}
\email{mario.bijelic@princeton.edu}
\affiliation{%
  \institution{Princeton University}
  \country{USA}
  }


  


\author{Johannes Froesch}
\email{jfroech@uw.edu}
\author{Arka Majumdar}
\affiliation{%
  \institution{University of Washington}
  \streetaddress{1410 NE Campus Pkwy}
  \city{Seattle}
  \state{Washington}
  \country{USA}
  \postcode{98195}
  }
\email{arka@uw.edu}

\author{Felix Heide}
\affiliation{%
  \institution{Princeton University}
  \streetaddress{35 Olden St}
  \city{Princeton}
  \state{New Jersey}
  \country{USA}
  \postcode{08540}
  }
  
\email{fheide@princeton.edu}






\begin{abstract}
Today's commodity camera systems rely on compound optics to map light originating from the scene to positions on the sensor where it gets recorded as an image. To record images without optical aberrations, i.e., deviations from Gauss' linear model of optics, typical lens systems introduce increasingly complex stacks of optical elements which are responsible for the height of existing commodity cameras. In this work, we investigate \emph{flat nanophotonic computational cameras} as an alternative that employs an array of skewed lenslets and a learned reconstruction approach. The optical array is embedded on a metasurface that, at 700~nm height, is flat and sits on the sensor cover glass at 2.5~mm focal distance from the sensor. To tackle the highly chromatic response of a metasurface and design the array over the entire sensor, we propose a differentiable optimization method that continuously samples over the visible spectrum and factorizes the optical modulation for different incident fields into individual lenses. We reconstruct a megapixel image from our flat imager with a \emph{learned probabilistic reconstruction} method that employs a generative diffusion model to sample an implicit prior. To tackle \emph{scene-dependent aberrations in broadband}, we propose a method for acquiring paired captured training data in varying illumination conditions. We assess the proposed flat camera design in simulation and with an experimental prototype, validating that the method is capable of recovering images from diverse scenes in broadband with a single nanophotonic layer.
 \end{abstract}

\begin{teaserfigure}  
  \includegraphics[width=1.0\linewidth]{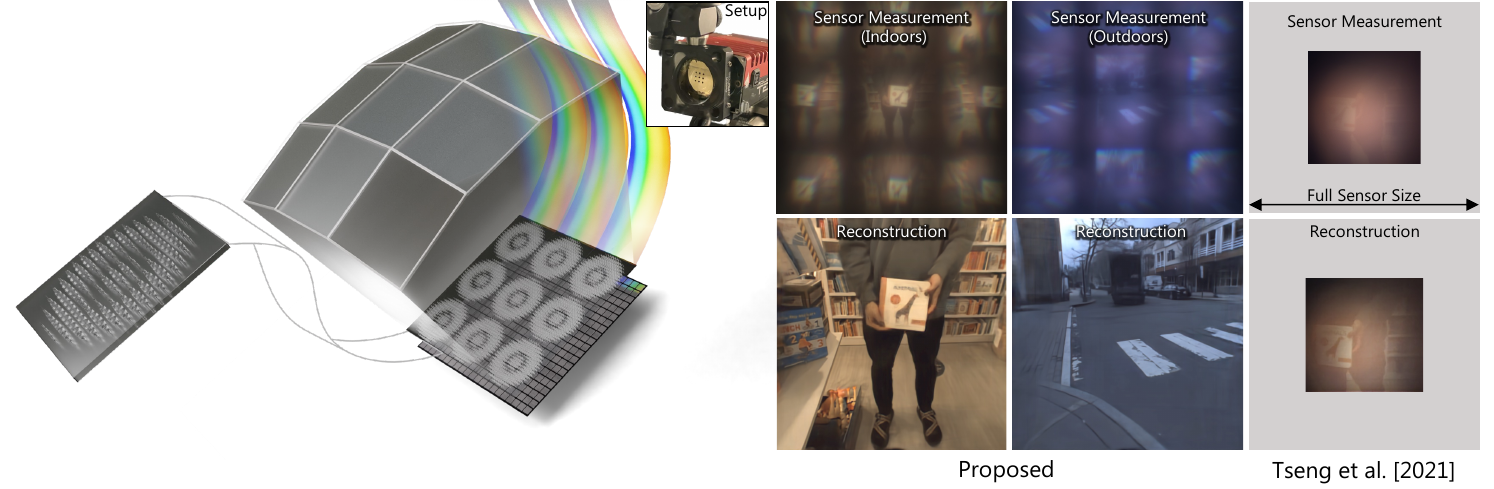}\vspace*{-8pt}
\caption{We propose a thin nanophotonic imager that employs a learned array of metalenses to capture a scene in-the-wild. We devise an array of lenses mounted directly on top of the sensor cover glass (top center). Each lens in our array (illustrated on the left) is a flat (700~nm thick) metasurface area of nano-antennas which we design to focus light across the visible spectrum. The peripheral elements capture images at slanted field angles, making it possible to capture a wide field of view of 100$^\circ$, more than twice as large as the most similar design from Tseng et al.~\shortcite{neural_nanooptics} shown on the right side. The proposed computational camera is capable of recovering images outside of lab conditions under broadband illumination. We illustrate here the matching physical image size as measured on the sensor, with the gray area on right illustrating same sensor size.}
  \label{fig:teaser}
\end{teaserfigure}

\maketitle


\section{Introduction}
\label{sec:intro}
Cameras have become a ubiquitous interface between the real world and computers with applications across domains in fundamental science, robotics, health, and communication. Although their applications are diverse, \emph{today's cameras acquire information in the same way they did in the 19th century}: they focus light on a sensing plane using a stack of lenses that minimize deviations from Gauss's linear model of optics~\cite{gauss1843dioptrische}. In this paradigm, increasingly complex and growing sets of lenses are designed to record an image. Since the microfabrication revolution in the last century brought us miniaturized sensors and electronic chips, it is now these optical systems that dictate a camera's size and weight and prohibit miniaturization without drastic loss of image quality~\cite{peng2016diffractive, stork2014optical, asif2016flatcam}. For example, the optical stack of the iPhone 13 contains more than seven elements that make up the entire 8~mm of the camera length responsible for the camera bump. Unfortunately, attempts to use thinner single-element optics~\cite{venkataraman2013picam, peng2016diffractive, stork2014optical}, amplitude masks close to the sensor~\cite{asif2016flatcam,khan2020flatnet}, or diffusers~\cite{kuo2017diffusercam,antipa2018diffusercam} instead of focusing optics have not been able to achieve the high image quality that conventional compound lens systems deliver. 

The emerging field of nanophotonic metaoptics suggests an alternative. These optical devices rely on quasi-periodic arrays of sub-wavelength scatterers that are engineered to manipulate wavefronts. \emph{In principle,} this approach promises new capabilities to drastically reduce the size and weight of these elements. The unprecedented ability to engineer each nanoscatterer enables optical functionality that is extremely difficult, if not impossible, to achieve using conventional optics: spectral and spatial filters~\cite{camayd2020multifunctional}, polarization imagers~\cite{arbabi2018full}, compact hyperspectral imagers~\cite{faraji2019hyperspectral}, depth sensors~\cite{colburn2020metasurface}, and even image processors~\cite{zhou2020flat}. 
Moreover, these flat optical elements are ultrathin, with a device thickness around an optical wavelength. 

The imaging performance of existing meta-optics, however, is far from that of their refractive counterparts lenses~\cite{colburn2020metasurface,neural_nanooptics}. While these lenses are corrected for wavelengths across the visible regime, the image quality is not on par with refractive lenses: third-order Seidel aberrations (e.g., coma, field curvature, and distortion) remain uncorrected as they are not even considered in the design procedure for these devices. Furthermore, the small apertures of the metalenses used ($\approx$ 50--100$\mu$m) severely limit the achievable angular resolution and total light collection (reducing the signal-to-noise ratio).

Increasing the field of view and aperture of these metalenses while simultaneously maintaining and improving aberration correction faces fundamental challenges: metasurfaces are inherently chromatic like any diffractive optics. For a metalens designed for a specific wavelength, the positions of the rings of constant phase decides the lens focusing behavior. When the incident wavelength changes, however, the imparted phase exhibits erroneous phase-wrapping discontinuities that vary significantly from the ideal response expected for the incident, non-design wavelengths ~\cite{arbabi2016multiwavelength}; this is the primary reason why metasurfaces exhibit chromatic aberrations. Recently, dispersion engineering has been investigated to design a metasurface to uniformly focus light across the full visible wavelength range ~\cite{wang2018broadband,chen2018broadband}. This technique relies on designing scatterers with not only a desired phase but also its higher-order response in the form of group delay and group delay dispersion. Recent work finds that there is a fundamental limit on the optical bandwidth for a dispersion-engineered metalens given a feasible aspect ratio and, therefore, small lens thickness -- a limit that arises from the inherent time-bandwidth product~\cite{presutti2020focusing}. The most successful approach to broadband imaging with metasurface optics from Tseng et al.~\cite{neural_nanooptics} relies on end-to-end computational imaging and jointly designs lens parameters and computational reconstruction~\cite{sitzmann2018end,LearnedLargeFovImaging2019} with a differentiable forward model. Despite achieving increased image quality, the design limitations and computational and memory consumption of this approach is also fundamentally limited to designing \emph{nanophotonic optics with a limited field of view of 40$^\circ$, optimized for narrow wavelength bands, and low image resolutions of a few kilopixels}~\cite{neural_nanooptics}. We aim to tackle this issue and design broadband computational nanophotonic cameras that can lift this limitation and make thin cameras possible, more than two orders of magnitude thinner and lighter than today. 

In this work, we propose a flat camera that relies on an array of nanophotonic optics, which are learned for the broadband spectrum, and a computational reconstruction module that recovers a single megapixel image from the array measurements. The camera employs a \emph{single flat optical layer sitting on top of the sensor cover glass} at 2.5~mm focal distance from the sensor. We introduce a differentiable forward model that approximates the highly chromatic wavefront response of a metasurface atom conditioned on the structure parameters in a local region. Instead of full-wave simulation methods that do not allow for simulating apertures larger than tens of microns across the visible band due to prohibitive memory and computational requirements, this differentiable model allows us to piggy-back on distributed machine learning methods and learn nanophotonic imaging across the entire band by stochastic gradient optimization over the continuous spectrum -- in contrast to Tseng et al.~\cite{neural_nanooptics} who optimize over the three fixed wavelengths of an OLED display. We achieve high-quality imaging performance across the entire visible band and more than double the field of view of existing approaches to 100$^\circ$ by separating the optical modulation for different optical fields into individual lenses in an array. We recover a latent image from our flat imager with a learned optimization method that relies on a diffusion model as a natural image prior. To tackle reconstruction in broadband illumination, we introduce a novel method to capture large datasets of paired ground-truth data in real-world illumination conditions. 

Specifically, we make the following contributions: 
\begin{itemize}
	\item We introduce a flat on-sensor nanophotonic array lens that decomposes the joint optimization over field angle and broadband focusing into several subproblems with several smaller field of view. We propose a stochastic optimization method for designing the decomposed broadband array elements.  
	\item We propose a novel learned probabilistic reconstruction method that relies on the physical forward model combined with a learned diffusion model as prior. To train the method, we propose an approach to capture paired real-world datasets.
	\item We analyze our method in simulation and compare the proposed method to alternative flat optical systems.
	\item We assess the method with a prototype camera system and compare it against existing metasurface designs. We confirm that the method achieves favorable image quality compared to existing metasurface optics across the entire spectrum and with a large field of view with a flat optical system on the sensor cover glass.
\end{itemize}

We will release all code, optical design files, and datasets.
 
\paragraph{Limitations}

Compared to traditional cameras with larger optical systems, the proposed flat camera shares with existing computational flat cameras the need for GPU processing with high power consumption. Despite this limitation, the compute resources on modern smartphones present opportunities for the efficient implementation of the proposed reconstruction method on custom ASICs, potentially enabling fast inference on edge devices in the future. 
\rev{Our prototype does not use the full available optical aperture. To avoid optical baffles and overlap, we space out the sublenses in the array over non-contiguous regions, resulting in low total light efficiency. We also do not explicitly consider fabrication inaccuracies.}

\section{Related Work}
\label{sec:related}
\paragraph{Flat Computational Cameras.}
Researchers have investigated several directions to reduce the height and complexity of existing compound camera optics. A line of work aims at reducing a complex optical stack of a handful to a dozen elements, to a single refractive element~\cite{heide2013high,li2021universal,Schuler2013AML,tanida2001thin} resulting in geometric and chromatic aberrations. Trading optical for computational complexity to address the introduced aberrations, these approaches have achieved impressive image quality comparable to a low-resolution point-and-shoot camera. Venkataraman et al.~\shortcite{venkataraman2013picam} suppress chromatic aberrations by using an on-sensor array of color-filtered single lens elements, which turns the deconvolution problem into a chromatic light field reconstruction approach that is challenging to solve without artifacts. 
All proposed single-element refractive and diffractive cameras~\cite{heide2016encoded,Peng2015ComputationalIU,Peng2016DiffractiveAchromat} have in common that, although the optical stack itself decreases in height (less than a micron for diffractive elements), they require \emph{long backfocal distances} of more than $\SI{10}{mm}$ prohibiting thin cameras. 
Lensless cameras\cite{White2020ASP,asif2016flatcam,khan2020flatnet,antipa2018diffusercam,liu2019single_diffuser,monakhova2020spectraldiffuser,kuo2017diffusercam} instead replace the entire optical stack with amplitude masks or diffusers that scramble the incoming wavefronts. Although this approach allows for thin cameras of a few millimeters in height, the information of a given scene point is distributed over the entire sensor.
The light efficiency of these cameras is half of that of conventional lens systems, and recovering high-quality images from the coded measurements with large point spread functions of global support is challenging and, as such, the ill-posedness of the underlying reconstruction problem severely limits spatial resolution and requires long acquisition times. 
Using diffusers as caustic lenses has been investigated for 2D photography~\cite{kuo2017diffusercam}, 3D imaging~\cite{antipa2018diffusercam} and microscopy~\cite{kuo2020chip}. In addition to resulting in a challenging ill-posed reconstruction problem, the optimal distance from the diffuser to the sensor may vary from one diffuser to another~\cite{boominathan2020phlatcam}. In this work, we investigate an array of steered metasurface lenses as an alternative that allows for a short backfocal distance without mandating aberrations with global support or reducing light efficiency. 

\begin{table}[ht]
    \setlength{\tabcolsep}{0em}
    \centering
    \footnotesize
    \caption{
        Comparison of related work on thin cameras, where each criterion is fully~\greencheck, partially~\yellowcheck, or not~\redcheck met.
        See text for discussion.
        }
    \begin{tabularx}{\linewidth}{m{0.24\linewidth}XXXXXX}
    \toprule
    &
    {\footnotesize FlatCam \shortcite{asif2016flatcam}} &
    {\footnotesize DiffuserCam \shortcite{kuo2017diffusercam}} &
    {\footnotesize PiCam \shortcite{venkataraman2013picam}} &
    {\footnotesize Peng et al.~\shortcite{Peng2016DiffractiveAchromat}} &
    {\footnotesize Tseng et al.~\shortcite{neural_nanooptics}} &
    {\footnotesize Ours} \\
    \midrule
    \multicolumn{6}{l}{\textbf{Camera Characteristics}}\\ 
    \midrule
    On-Sensor (< 2mm) & 
    \green & 
    \yellow & 
    \green & 
    \red & 
    \green &
    \green \\
    Light Efficiency & 
    \red & 
    \yellow & 
    \yellow & 
    \green & 
    \red &
    \green \\
    Broadband & 
    \green & 
    \green & 
    \green & 
    \yellow & 
    \yellow &
    \green \\
    Wide Field of View & 
    \yellow & 
    \yellow & 
    \yellow & 
    \red & 
    \yellow &
    \green \\
    MTF & 
    \red & 
    \red & 
    \yellow &
    \green & 
    \red  &
    \green \\
    Fabrication & 
    \green & 
    \yellow & 
    \red & 
    \red & 
    \green & 
    \green \\
    Well-posedness & 
    \red & 
    \red & 
    \yellow &
    \yellow & 
    \yellow  &
    \green \\
    \bottomrule
\end{tabularx}
    \label{tab:related_work}
\end{table}

\paragraph{Metasurface Optics.}

Over the last few years, recent advancements in nanofabrication have made it possible for researchers to investigate optics by using quasi-periodic arrays of subwavelength scatterers to modify incident electromagnetic radiation. These ultra-thin metasurfaces allow the fabrication of freeform surfaces using single stage lithography. Specifically, meta-optics can be fabricated by piggy-backing on existing chip fabrication processes, such as deep ultraviolet lithography (DUV), without error-prone multiple etching steps required for conventional diffractive optical elements~\cite{Shi2022SeeThroughObstructions}. Each scatterer in a metasurface can be independently tailored to modify amplitude, phase, and polarization of wavefronts -- light can be modulated with greater design freedom compared to conventional diffractive optical elements (DOEs)~\cite{Engelberg2020TheAO,Lin298Dielectric,Mait2020PotentialAppl,Peng2019LearnedLF}. With these theoretical advantages in mind, researchers have investigated flat meta-optics for imaging~\cite{Colburn2018MetasurfaceOF,Yu2014FlatOW,Aieta2012AberrationFree,Lin2021EndtoEnd}, polarization control~\cite{Arbabi2015DielectricMF}, and holography~\cite{Zheng2015MetasurfaceHR}. However, existing meta-optics suffer from severe chromatic and geometric aberrations making \emph{broadband imaging outside the lab infeasible with existing designs}. In contrast to diffractive optics, the wavelength-dependent aberrations are a direct result from non-linear imparted phase~\cite{Yu2014FlatOW,Lin298Dielectric,Aieta2015MultiwavelengthAM,Wang2018ABA}. While methods using dispersion engineering~\cite{Ndao2020OctaveBP,Shrestha2018BroadbandAD,Arbabi2017ControllingTS,khorasaninejad2017achromatic,wang2017broadband} are successful in reducing chromatic aberrations, these methods are limited to aperture sizes of tens of microns~\cite{Presutti2020FocusingLimits}. Most recently, Tseng et al.~\cite{neural_nanooptics} have proposed an end-to-end differentiable design approach for meta-optics that achieves full-color image quality with a large 0.5~mm aperture. However, while successful in imaging tri-chromatic bands of an OLED screen, their method does not perform well outside the lab and suffers from severe blur for fields beyond 40$^\circ$. Recent advanced nano fabrication techniques have also made compact conventional cameras with wafer-level compound optics possible, e.g., OVT CameraCube \footnote{https://www.ovt.com/technologies/cameracubechip/} which, however, offers limited resolution and FoV. The proposed array design in this work optimizes image quality over the full broadband spectrum across the 100$^\circ$ deg FoV without increasing the backfocal length. Our method can potentially allow for one-step fabrication of the metalens directly on the camera sensor coverglass in the future, further shrinking existing wafer-level multi-element compound lens camera designs.

\paragraph{Differentiable Optics Design.}
Conventional imaging systems are typically designed in a sequential approach, where the lens and sensors are hand-engineered concerning specific metrics such as RMS spot size or dynamic range, independently of the downstream camera task. Departing from this conventional design approach, a large body of work in computational imaging has explored jointly optimizing the optics and reconstruction algorithms, with successful applications in color image restoration~\cite{Chakrabarti2016LearningSM,peng2019learned}, microscopy~\cite{Horstmeyer2017ConvolutionalNN,kellman2019data,Nehme2019DenseTD,Shechtman2016MulticolourLM}, monocular depth imaging~\cite{Chang2019DeepOF,Haim2018DepthEF,He2018LearningDF,Wu2019PhaseCam3DL}, super-resolution and extended depth of field~\cite{sitzmann2018end,sun2021end}, time-of-flight imaging~\cite{Marco2017DeepToFOR,Su2018DeepET,chugunov2021mask}, high-dynamic range imaging~\cite{Sun_2020_LearnedOpticHDR,metzler2019deep}, active-stereo imaging~\cite{baek2021polka}, hyperspectral imaging~\cite{baek2021single}, and computer vision tasks~\cite{Tseng2021DeepCompoundOptics}.
In this work, we take a hybrid approach wherein we first optimize a nanophotonic lens array camera, designed with an inverse filter as an efficient proxy for the reconstruction method. We then devise a novel probabilistic deconvolution method conditioned on the measured signals for full-color image restoration, computationally compensating residual aberrations. 

\section{Nanophotonic Array Camera}
\label{sec:method}

\begin{figure}[t!]
	\centering
		\includegraphics[width=\columnwidth]{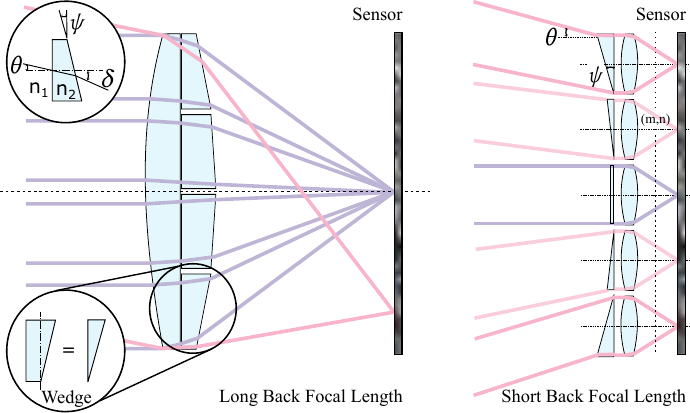}
		\caption{Lens as a Combination of Prisms. A lens refracts an incoming set of parallel rays and bends them towards a focal point. With a piece-wise linear approximation, a lens can be thought of as stacked local, infinitesimally small, wedge prisms. We depart from a single large lens with long back focal length, re-organize the prisms, and represent a large lens as a combination of appropriately chosen wedges and lenses with short focal length for high-quality camera imaging.
}
		\label{fig:prism_lens}
\end{figure}

\begin{figure}[t!]
	\centering
		\includegraphics[width=\columnwidth]{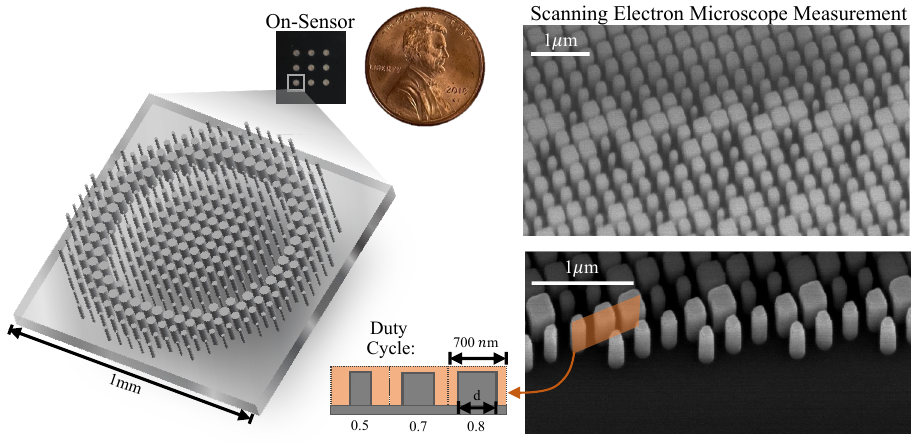}
		\caption{Ultra-thin and Compact Metalens Array. Our on-sensor imaging optic, which is smaller than a penny, consists of an array of metalenses as shown above. Each metalens is made of optimized nano-antennas of size \SI{350}{nm}, significantly smaller than the wavelength of visible light. Our optimized metalens nano-antenna structures scatter light from the entire visible spectrum.}
		\label{fig:metasurface}
\end{figure}

In this section, we describe the nanophotonic array camera for thin on-sensor imaging. We design the imaging optic by learning an array of short back focal length metalenses with carefully designed phase profiles, enabling the camera to capture a scene with a large viewing angle. A learned image reconstruction method recovers the latent image from the nanophotonic array camera resulting in a thin on-chip imaging system. In the following, we first describe the nanophotonic array optic. In the remainder of this section, we then derive the differentiable image formation model for the metalens array which we rely on to learn the phase profiles. In Sec.~\ref{sec:image_reconstruction}, we describe our reconstruction method. 

\subsection{Lens as a Combination of Prisms}
A lens can be thought of as analogous to a series of \emph{continuous} prisms as shown in Figure~\ref{fig:prism_lens}. 
A conventional prism refracts light causing its path to bend as
\begin{equation}
    \delta = \theta - \psi + \sin^{-1}\left(\sqrt{n^2 - \sin^2\theta}\sin\psi - \sin\theta \cos\psi \right),
\end{equation}
where $n=n_2/n_1$ is the relative refractive index, $\psi$ is the wedge angle of the prism, $\theta$ is the angle of incident light and $\delta$ is the angle of deviation, illustrated in Fig.~\ref{fig:prism_lens}. 
The greater the wedge angle, the greater the deviation of light path is. For smaller wedge and incident angles, the angle of deviation can be approximated as
\begin{equation}
    \delta \approx (n-1)\psi .
\end{equation}
Therefore, narrow-angle prisms in an imaging setup merely result in a shift in image position.
Specifically, light passing through the optical center of a lens where the surfaces are parallel to each other yields no prismatic effect in the center. However, light passing through the periphery of a lens experiences prismatic effects. The increasing angle between the opposing surfaces further from the lens center causes light to bend more and more, allowing the lens to focus light. 
Therefore, stacking a series of tiny prisms effectively makes a lens, notwithstanding the presence of aberrations.
However, since different wavelengths are refracted differently, a large angle prism also causes different wavelengths to spread out in the image, causing chromatic aberrations due to the spectrum of colors from white light, coma, and astigmatism. 

In this work, we design our optical layer as a combination of tiny co-optimized lens and prism phase elements, illustrated in Fig.~\ref{fig:prism_lens}, where the prisms help expand the field of view and the lenses reduce the back focal length. This allows us to devise a nanophotonic lens array to approximate the large lens, but trading off focal length with aberrations. 
Moreover, we also optimize the lens to correct some of the dispersion-based aberrations caused by the prism phases.

\subsection{Radially Symmetric Nanophotonic Array}
Nanophotonic meta-optics are ultrathin optical elements that utilize subwavelength nano-antenna scatterers to modulate incident light. Typically, these nano-antenna structures are designed for modulating the phase of incident light at a single nominal design wavelength, making meta-optics efficient for monochromatic light propagation. However, we require the meta-optic to achieve the desired phase modulation at all visible wavelengths to design a broadband imaging lens. 

\paragraph{Metasurface Design Space}

We design metasurfaces that consist of silicon nitride nanoposts with a height of 700 nm and a pitch of 350 nm on top of a fused silica (n = 1:5), see Fig.~\ref{fig:metasurface}. The rectangular pillars are made of a high refractive index material that is transparent. We keep these parameters fixed and then optimize the width (= length) of the nano-antennas between 100--300 nm, determining what we call the local ``duty cycle'', see again Fig.~\ref{fig:metasurface}.

In a local neighborhood of these nano-antennas, we are able to simulate the phase for a given duty cycle using rigorous-coupled wave analysis (RCWA), which is a Fourier-domain method that solves Maxwell's equations efficiently for periodic dielectric structures. As such, in the following, we characterize metalenses with their local phase, which we tie to the structure parameters, i.e., the duty cycle, via a differentiable model.

\paragraph{Radially Symmetric Metalens Array}
We model the metasurface phase $\phi$, which we treat as a differentiable variable in our design, as a radially symmetric per-pixel basis function
\begin{equation}
    \phi(x_i,y_j) = \phi(r), \hspace{2mm} r = \sqrt{x_i^2 + y_j^2}, \hspace{1mm} i,j \in \{1,2,...,N\},
\label{eq:radially-symmetric-phase}
\end{equation}
where $N$ is the total number of pixels along each axis, $(x_i, y_j)$ denotes the nano-antenna position and $r$ is its distance from the optical axis of the metalens. The per-pixel basis allows each nano-antenna along one radius of the metasurface to vary independently of the other nano-antennas without constraints. We constrain the metalens to be radially symmetric as opposed to optimizing the phase in a per-pixel manner to avoid local minima. Additionally, a spatially symmetric design imparts a spatially symmetric PSF which reduces the computational burden as it allows the simulation of the full field-of-view by only simulating PSFs along one axis. 

We impose an additional wedge phase of varying wedge angles over each metalens element to achieve a wider field of view. Therefore, for an $M\times N$ nanophotonic array, the phase of each element is given by
\begin{equation}
    \phi^{m,n}(x_i, y_j) = \phi^{m,n}(r) = \phi(r) + 
    \frac{2\pi}{\lambda_w} \left(x_i\sin\psi_\mathrm{x}^{m,n} + y_i\sin\psi_\mathrm{y}^{m,n} \right),
\label{eq:array-phase}
\end{equation}
where $\phi^{m,n} (x_i, y_j)$ is the phase modulation at the $(x_i, y_j)$-th nano-antenna of metalens in $m$-th row and $n$-th column, $\lambda_w$ is the wavelength for which the wedge phase was defined, and $(\psi_x, \psi_y)$ are the selected wedge angles along each axis. Note that for given wedge angles of a metalens element in the array, the additional wedge phase is constant whereas the radially symmetric phase is optimizable. 

Since the phase is defined only for a single nominal design wavelength, we apply two operations in sequence at each scatterer position in our metasurface: 1) a phase-to-structure inverse mapping to compute the scatterer geometry at the design wavelength for a given phase and 2) a structure-to-phase forward mapping to calculate the phase at other target wavelengths given a scatterer geometry. To allow for direct optimization of the metasurface phase, we model both the above operators as polynomials to ensure differentiability, which we describe below. 

\paragraph{RCWA Proxy Mapping Operators}
We describe the scatterer geometry with the duty cycle of nano-antennas and analyze its modulation properties using rigorous coupled-wave analysis (RCWA). The phase as a function of duty cycle of the nano-antennas must be injective to achieve a differentiable mapping from phase to duty cycle. 
To this end, we fit the phase data of the metalens at the nominal design wavelength to a polynomial proxy function of the form
\begin{equation}
d(r) = \sum_{i=0}^N a_i \left(\frac{\phi(r)} {2\pi}\right)^{2i}, 
\label{eq:phase-to-duty}
\end{equation}
where $d(r)$ is the required duty cycle at a position $r$ from the optical axis on the metasurface, $\phi(r)$ is the desired phase for the nominal wavelength $\lambda_0$, and the parameters $a_i$ are fitted. We set the nominal wavelength $\lambda_0 = \SI{452}{nm}$ for all of our experiments. 

After applying the above phase-to-scatterer inverse mapping to determine the required physical structure, we compute the resulting phase from the given scatterer geometry for other wavelengths using a second scatterer-to-phase proxy function. This forward mapping function maps a combination of the nano-antenna duty cycle and incident wavelength to an imparted phase delay. 
We model this proxy function by fitting the pre-computed transmission coefficient of scatterers under an effective index approximation~\cite{neural_nanooptics} to a radially symmetric second-order polynomial function of the form
\begin{equation}
\widetilde{\phi}(r, \lambda) = \sum_{n=0}^2 \sum_{m=0}^2 b_{nm} d(r)^{n} \lambda^{m}, \hspace{2mm} n+m \leq 2,
\label{eq:duty-to-phase}
\end{equation}
where $\lambda$ is a \emph{non-nominal} wavelength. 
Specifically, we compute the transmission coefficient data $\mathcal{C}_\textsc{meta}$ using RCWA and then fit the polynomial to the underlying RCWA-computed transmission coefficient data using linear least squares. For details on fitted polynomial coefficients for the inverse and forward mappings and additional details on proxy functions, please see Supplementary Material.

\paragraph{Single Lens Element Image Formation}
With the metalens phase described by Eq.~\eqref{eq:array-phase} and the mapping operators defined in Eq.~\eqref{eq:phase-to-duty} and Eq.~\eqref{eq:duty-to-phase}, we compute the phase modulation for a broadband incident light. Using a fast Fourier transform (FFT) based band-limited angular spectrum method (ASM), we calculate the PSFs produced by each metalens in the array as a function of wavelength and field angle to model full-color image formation over the entire field of view. The spatially varying PSF as produced by each element in the nanophotonic array for an incident beam of wavelength $\lambda$ at an angle $\theta$ is
\begin{equation}\label{eq:forward_model}
    \mathrm{PSF_{\theta, \lambda}^{m,n}} = f_\textsc{meta}\big( \phi^{m,n}(r), \theta, \textsc{C}_\textsc{meta} \big),
\end{equation}
where $\phi^{m,n}(r)$ is the optimizable radially symmetric metasurface phase and $\textsc{C}_\textsc{meta}$ are the set of fixed parameters such as aperture and focal length of the metalens, and $f_\textsc{meta}(\cdot)$ is the angular spectrum method as a propagation function that generates the PSF $\mathbf{k}$ for a given metasurface phase. Finally, the RGB image on the sensor plane is 
\begin{equation}
    \mathbf{S} = \mathbf{I} \otimes \mathbf{k} + \eta_\textsc{sensor},
    \label{eq:sensor-measurement}
\end{equation}
where $\otimes$ is a convolution operator, $\mathbf{I}$ is the groundtruth RGB image, and $\eta_s$ is the sensor noise modeled as per-pixel Gaussian-Poisson noise.

Specifically, for an input signal $x \in [0,1]$ at a sensor pixel location, the measured noisy signal $f_\textsc{sensor}(x)$ is given by
\begin{equation}\label{eq:sensor_noise}
f_\textsc{sensor}(x) = \eta_g(x, \sigma_g) + \eta_p(x, a_p),
\end{equation}
where $\eta_g(x, \sigma_g) \sim \mathcal{N}(x, \sigma_g^2)$ is the Gaussian noise component and $\eta_p(x, a_p) \sim \mathcal{P}(x / a_p)$ is the Poisson noise component. 

\paragraph{Spatially Varying Array Image Formation}
We simulate the spatially varying aberrations in a patch-wise manner. We first divide the overall FoV into an $M \times N$ grid of patches for a nanophotonic array with $M \times N$ metalens elements. For incident broadband light at field angle $\theta$, we then compute $\textsc{PSF}_{\theta, \lambda}$ for each metalens element in the array with varying wedge angles, see Eq.~\ref{eq:array-phase}. While we use $\textsc{PSF}_{\theta, \lambda}$ for the image formation forward model, we permute the PSFs for different wavelengths for deconvolution. This process acts as a regularization to the PSF design and avoids variance across the spectrum, essential for robust imaging in the wild.  After design and fabrication, we account for mismatches between the simulated PSF by our proxy model and the experimentally measured PSF by performing a PSF calibration step. 

\paragraph{Differentiable Nanophotonic Array Design}
With a measurement $\mathbf{S}$ as input, we recover the latent image as
\begin{equation}
\Tilde{\mathbf{I}} = f_\textsc{deconv}(\mathbf{S}, \textbf{k}, \textsc{C}_\textsc{deconv}),
\end{equation}
where $\textsc{C}_\textsc{deconv}$ are the fixed parameters of the deconvolution method. To make our lens design process efficient, we employ an inverse filtering method in the design of our optic which does not require training and allows it to be computed in one step, as opposed to the proposed method in Sec.~\ref{sec:image_reconstruction}, see Supplemental Material for detail.

With this synthetic image formation model in hand, our nanophotonic array imaging pipeline allows us to apply first-order stochastic gradient optimization to optimize for the metalens phases that minimize the error between the ground truth and recovered images. In our case, given an input RGB image $\mathbf{I}$, we aim to find a metalens array that will recover $\mathbf{I}$ with high fidelity with short back focal length to achieve compact and ultra-thin imaging device with wide FoV. 
To design our optical system, we minimize the per-pixel mean squared error and maximize the perceptual image quality between the target image $\mathbf{I}$ and the recovered image $\Tilde{\mathbf{I}}$.
To this end, we use first-order stochastic gradient descent solvers to optimize for individual metalens elements in the nanophotonic array as follows
\begin{equation}
\Tilde{\phi}(r)^{m,n} = \argmin{\phi} \sum_{i=1}^T \sum_{\theta, \lambda} \mathcal{L} \left(\Tilde{\mathbf{I}}_{\theta, \lambda}^{(i)} , \mathbf{I}_{\theta, \lambda}^{(i)}\right),
\label{eq:objective-func}
\end{equation}
where $T$ is the total number of training image samples and the images are measured by the $(m,n)$-th metalens in the array and the loss function
\begin{equation}
    \mathcal{L} = \mathcal{L}_\textsc{mse} + \mathcal{L}_\textsc{lpips}.
\end{equation}
Specifically, we design the metalens to work in the entire broadband visible wavelength range and modulate the incident wave fields over a $60^\circ$ FoV. We notice that PSFs vary smoothly across the FoV and hence we sample it in regular intervals of $15^\circ$ during optimization, whereas the wavelengths are sampled in intervals of \SI{50}{nm} over the visible range. We use the Adam optimizer with a learning rate of $0.001$ running for 15 hours over the dataset described in Sec.~\ref{sec:dataset} to optimize for the meta-optic phase.  
We further fine-tune the metalens phase to suppress side lobes in the PSF to eliminate the haze that corrupts the sensor measurements, especially the ones captured in the wild. Once the optimization is complete, we use the optimized radially symmetric metalens with the appropriate wedge phases to manufacture our meta-optic.

\paragraph{Full Spectrum Phase Initialization}
To aid the optimization from above, we propose a full spectrum metalens phase initialization wherein a rotationally symmetric metalens phase is optimized to maximize the focal intensity at the center of the field of view. Specifically, we initialize the optimization described in Eq.~\ref{eq:objective-func} with the solution to another optimization problem with the following objective
\begin{equation}
    \Tilde{\phi}(r) = \argmin \phi \sum_{\lambda=400}^{700}  -f_\textsc{meta} ( \phi, \theta = 0,\textsc{C}_\textsc{meta})\Big|_{(x_s, y_s) = (0,0)}.
\end{equation}
where $(x_s, y_s)$ are the coordinates on the sensor plane.
In other words, the solution to the above optimization problem finds a metalens phase that focuses all broadband light energy at the center of the sensor plane, thereby significantly reducing chromatic artifacts. We sample the wavelengths in steps of \SI{10}{nm} and further use a per-pixel error function on the computed PSF in order to further improve the phase initialization. Note that similar to the phase described in Eq.~\eqref{eq:radially-symmetric-phase}, we use a per-pixel basis for solving the above metasurface phase which we later use to initialize Eq.~\eqref{eq:objective-func}. 

Finally, the phase obtained by solving the optimization problem described in Eq.~\eqref{eq:objective-func} is fabricated and installed on the sensor of the prototype camera, see Sec.~\ref{sec:prototype}. The measurements by this ultra-thin compact camera follow Eq.~\eqref{eq:sensor-measurement}, and we next describe how the latent images are recovered.

\section{Probabilistic Image Recovery} 
\label{sec:image_reconstruction}

\begin{figure*}[t!]
\vspace*{-5pt}
	\centering
    \includegraphics[width=1.035\linewidth]{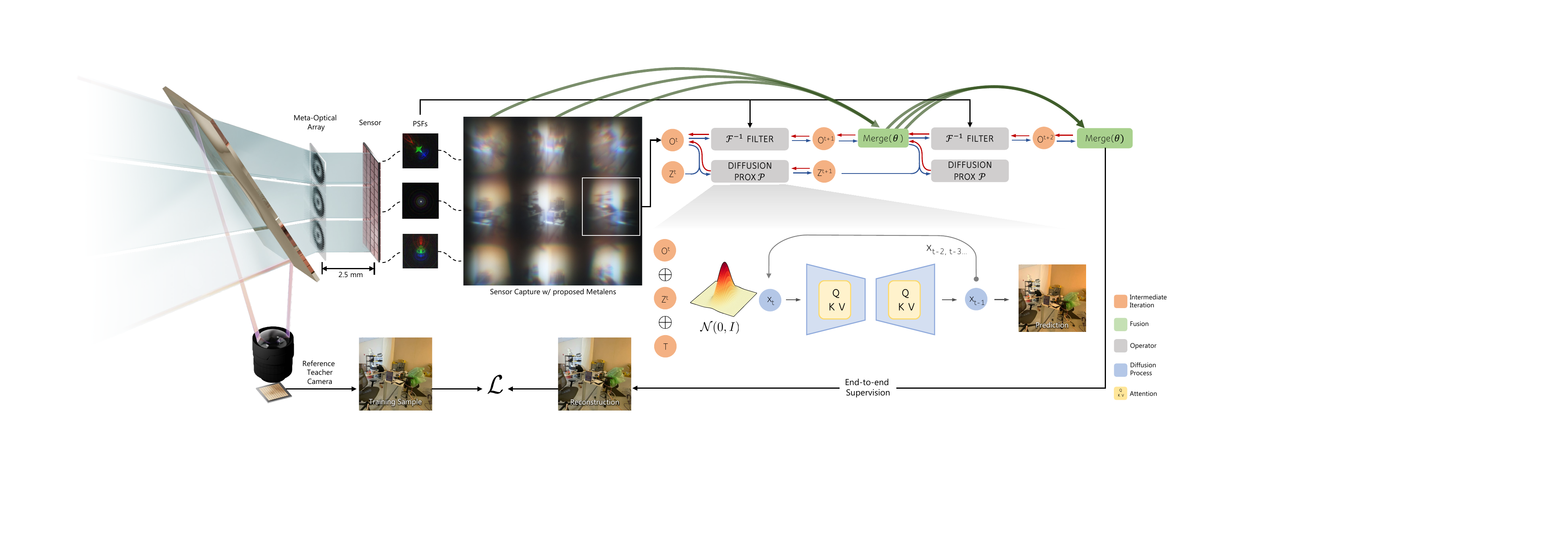}\vspace*{-5pt}
		\caption{Overview of Probabilistic Image Reconstruction. We propose a deconvolution method that relies on the physics-based forward model (PSFs illustrated on the left) along with a learned probabilistic prior (diffusion model on the bottom right). The proposed method is an unrolled optimization method that alternates between inverse filtering steps using the calibrated PSFs, diffusion steps that sample from the natural image manifold with conditioning on the current iteration, and a merging step that combines the image estimates from all sub-apertures (green). The unrolled optimization method is trained in an end-to-end fashion with paired training samples captured with a co-axial reference camera (bottom and left), see text for details.}
		\label{fig:framework}
\end{figure*}

This section describes how we recover images from measurements of the on-sensor array camera. We first formulate the image recovery task as a model-based inverse optimization problem with a probabilistic sampling stage that samples a learned prior. We solve the optimization problem via splitting and unrolling into a differentiable truncated solver. To learn a natural image prior along with the unrolled solver, we propose a probabilistic diffusion model that samples a multi-modal distribution of plausible latent images. For ease of notation, we first describe the image recovery algorithm for a single lens element before describing the recovery method for the entire array.

\subsection{Model-based Splitting Optimization for a Single Element}
\label{subsec:single-element-deconv}
We propose a method to recover the latent image $\mathbf{I}$ from the sensor measurement $\mathbf{S}$ that relies on the physical forward model described in Eq.~\eqref{eq:sensor-measurement}. We represent the spatially varying $\mathrm{PSF}$ of the array camera as $\mathbf{k}$ in the following for brevity. Following a large body of work on inverse problems in imaging~\cite{bertero2021introduction,venkatakrishnan2013plug,romano2017little}, we pose the deconvolution problem at hand as a Bayesian estimation problem. Specifically, we solve a maximum-a-posteriori estimation problem~\cite{laumont2022maximum} with an abstract natural image prior $\Gamma(\mathbf{I})$, that is
\begin{equation}
    \Tilde{\mathbf{I}} = \argmin{\mathbf{I}} \underbrace{\frac{1}{2} \big|\big| \mathbf{I}\otimes \mathbf{k} - \mathbf{S} \big|\big|^2}_{\text{Data Fidelity}} + \underbrace{\rho \Gamma(\mathbf{I})}_{\text{Prior Regularization}},
    \label{eq:min-problem}
\end{equation}
where $\rho > 0$ is a prior hyperparameter. However, instead of solving for the singular maximum of the posterior as a point estimate, we employ a \emph{probabilistic prior that samples} the posterior of all plausible natural image priors. In other words, this will allow us to sample multiple plausible reconstructions near the maximum.\newline

To solve Eq.~\eqref{eq:min-problem}, we split the non-linear and non-convex prior term from the  linear data fidelity term to result in two simpler subproblems via half-quadratic splitting. To this end, we introduce an auxiliary variable $\mathbf{z}$, and pose the above minimization problem as 
\begin{equation}
\begin{aligned}
    \argmin{\mathbf{I}} \frac{1}{2} \big|\big| \mathbf{I} \otimes \mathbf{k} - \mathbf{S} \big|\big|^2 + \rho \Gamma(\mathbf{z})
    \hspace{2mm} s.t. \hspace{2mm} \mathbf{z} = \mathbf{I}.
\end{aligned}
\label{eq:z}
\end{equation} 
We further reformulate the above minimization problem then as
\begin{equation}
    \argmin{\mathbf{I,z}} \frac{1}{2} \big|\big| \mathbf{I} \otimes \mathbf{k} - \mathbf{S} \big|\big|^2 + \rho \Gamma(\mathbf{z}) + \frac{\mu}{2}\big|\big| \mathbf{z} - \mathbf{I} \big|\big|^2,  
    \hspace{2mm} \mu \to \infty ,
    \label{eq:HQS-objective}
\end{equation}
where $\mu > 0$ is a penalty parameter, that $\mu \to \infty$ mandates equality $\mathbf{I} = \mathbf{z}$. We relax $\mu$ and solve the above Eq.~\eqref{eq:HQS-objective} iteratively by alternating between the following two steps,
\begin{equation}\label{eq:updates}
\begin{cases}
\mathbf{I}^{t+1} = \argmin{\mathbf{I}} \frac{1}{2} \big|\big| \mathbf{I} \otimes \mathbf{k} - \mathbf{S} \big|\big|^2 + \frac{\mu^t}{2}\big|\big| \mathbf{I} - \mathbf{z}^t \big|\big|^2, 
\\ 
\mathbf{z}^{t+1} = \argmin{\mathbf{z}} \frac{\mu^t}{2}\big|\big| \mathbf{z} - \mathbf{I}^{t+1} \big|\big|^2 + \rho \Gamma(\mathbf{z}).
\end{cases}
\end{equation}
where $t$ is the iteration index and $\mu^t$ is the updated weight in each iteration. We initialize our method with $\mu^0 = 0.1$ and exponentially increase its value for every iteration. Note that we solve for $\mathbf{I}$ given fixed values of $\mathbf{z}$ from the previous iteration and vice-versa.  

The first update from the iteration~\eqref{eq:updates} is a quadratic term that corresponds to the data term from Eq.~\eqref{eq:min-problem}. Assuming circular convolution, it can be solved in closed form with the following inverse filter update
\begin{equation}
    \mathbf{I}^{t+1} = \mathcal{F}^{\dag} \Bigg( \frac{\mathcal{F^*(\mathbf{k})}\mathcal{F}(\mathbf{S}) + \mu^t\mathcal{F}(\mathbf{I^t})}
    {\mathcal{F^*(\mathbf{k})} \mathcal{F(\mathbf{k})} + \mu^t} \Bigg) ,
    \label{eq:filter-step}
\end{equation}
where $\mathcal{F}(\cdot)$ denotes the Fast Fourier Transform (FFT), $\mathcal{F}^*(\cdot)$ denotes the complex conjugate of FFT, and $\mathcal{F}^{\dag}(\cdot)$ denotes the inverse FFT. Please refer to the Supplemental Material for the derivation.
\newline 

However, the second update from iteration~\eqref{eq:updates} includes the abstract regularizer and it is, in general, non-linear and non-convex. We learn the solution to this minimization problem with a diffusion model that allows us to probabilistically sample the solution space near the current iterate $\mathbf{I}^{t+1}$. Specifically we sample from a distribution $\Omega$ that is conditioned on the iterate $\mathbf{I}^{t+1}$ and the optimization penalty weights $\rho, \mu$ as inputs
\begin{equation}
    \mathbf{z}^{t+1} \sim \Omega(\mathbf{I}^{t+1}, \rho, \mu^t)
\end{equation}

Next, we describe how we learn and sample from this prior in our method.

\subsection{Diffusion-based Image Prior}
\label{subsec:diffusion}
We propose a diffusion-based prior $\Omega$~\cite{ddpm, sohl2015deep} to handle an ambiguity in deconvolution: multiple clear latent images can be projected to the same measurement $\mathbf{I}$. Diffusion provides a probabilistic approach to generate multiple samples, from which we can select the most suitable one. Figure~\ref{fig:framework} illustrates the proposed prior model. 

We first devise the forward process of diffusion by adding noise and learning to recover the clean image. We denote our input $x_0$ as $I^{gt}$, and our condition $c$ is defined as 
\begin{equation}
    c = I^{gt} \oplus \mathbf{S} \oplus  z^t \oplus  \mu^t \oplus  \gamma (T),
    \label{eq:ddpm_concat}
\end{equation}
where $I^{gt}$ is the ground truth latent image, $\mathbf{S}$ is the sensor measurement, $z^t$ is the auxiliary image coupling term defined in Eq. \eqref{eq:z}, $\mu^t$ is an update weight defined in Eq. \eqref{eq:HQS-objective}, and $\gamma (T)$ is a positional encoding of $T$ where $T \in [1,1000]$ is the timestep randomly sampled for each training iteration of the diffusion model. Note that the subscript $t$ in $z^t$ and $\mu^t$ refers to the HQS iteration from Eq.~\eqref{eq:updates}, separate from $T$ which refers to the diffusion timestep.

Here, $\oplus$ is the concatenation symbol, as we condition the inputs by concatenating them along the channel dimension and employ self-attention~\cite{attention} to learn corresponding features. 

To train our diffusion model, in each iteration we add Gaussian noise to the $I^{gt}$ of $x_0$ proportional to $T$ to obtain $x_t$. Specifically, we train the model $\Omega$ to recover $O^{gt}$ from $x_t$. Similar to \cite{chou2022diffusionsdf}, we recover $O^{gt}$ rather than the added noise. \rev{To tackle moderate misalignment in our dataset, we employ a Contextual Bilateral loss (CoBi) which is robust to misalignment of image pairs in both RGB and VGG-19 feature space\cite{zhang2019zoom}. Our overall training objective is }

\rev{
\begin{equation}
    \mathcal{L}_{\mathrm{diff}} = \mathcal{L}_{\mathrm{CoBi}}(\Omega(x_t), I^{gt})_{RGB} + \lambda \mathcal{L}_{\mathrm{CoBi}}(\Omega(x_t), I^{gt})_{VGG}
    \label{eq:diff_obj},
\end{equation}
}
\rev{where $\lambda$ is empirically selected via experimentation.} The architecture of our diffusion model is a UNet~\cite{Ronneberger2015UNetCN} following \cite{ddpm}. We provide further details in the Supplemental Document. 

During test time, our diffusion model performs generation iteratively. In the vanilla DDPM~\cite{ddpm}, generation is performed as follows
\begin{equation}
  z^\prime = (f \circ ... \circ f)( z_T,  T), \,\,
  f(x_t, t) = \Omega( x_t) + \sigma_t  \epsilon,
  \label{eq:ddpm_generation}
\end{equation}
where $z_T \sim \mathcal{N}(0,{I})$, $\sigma_t$ is the fixed standard deviation at the given timestep, and $\epsilon \sim \mathcal{N}(0,{I})$. However, this results in long sampling times. Instead, we follow DDIM~\cite{ddim}, and adopt a non-Markovian diffusion process to reduce the number of sampling steps. Furthermore, we use the ``consistency'' property that allows us to manipulate the initial latent variable to guide the generated output. As a result, $f(x_t, t)$ from Eq. \eqref{eq:ddpm_generation} can be defined as
\begin{equation} 
f(x_t, t) = \sqrt{\alpha_{t-1}}\left(\frac{{x}_t-\sqrt{1-\alpha_t} \Omega ({x}_t)}{\sqrt{\alpha_t}}\right)+ \sqrt{1-\alpha_{t-1}-\sigma_t^2} \cdot \Omega ({x}_t)+  \sigma_t  \epsilon.
\label{eq:ddim_generation}
\end{equation}
In practice, we find generation timesteps of 20 sufficient for our experiments. 

\subsection{Learned Array Deconvolution and Blending}
\label{subsec:array-deconv}
The nanophotonic array lens measures an array of images, each with a different FoV, which we deconvolve and merge together to form a wider FoV image. We employ the probabilistic image recovery approach described in Sec.~\ref{subsec:single-element-deconv} for deconvolving the array of images. Specifically, the individual latent images in the array are recovered by solving
\begin{equation}
    \Tilde{\mathbf{I}_{m,n}} = \argmin{\mathbf{I}_{m,n}} \frac{1}{2} \big|\big| \mathbf{I}_{m,n}\otimes \mathbf{k}_{m,n} - \mathbf{S}_{m,n} \big|\big|^2 + \rho \Gamma(\mathbf{I}_{m,n}),
    \label{eq:array-deconv}
\end{equation}
where $(m,n)$ corresponds to the sub-image in the $m$-th row and $n$-th column of the sensor array measurement. 
For solving this, we first acquire real PSF measurements $\mathbf{k}_{m,n}$ for each element of the metalens array. An sensor measurements $\mathbf{S}_{m,n}$ are acquired as a dataset of images captured in various indoor and outdoor environments, as described next in Sec.~\ref{sec:dataset}, to allow for learning the probabilistic prior $\Gamma$ over natural images.

The recovered array of latent images are finally blended together into a wider FoV super-resolved image to approximately match the sensor resolution. Given an $(m,n)$ array of input images $\{\mathbf{I}_{m,n} \}$ where $(m,n)=\{0,1,2,...\}$, our goal is to produce a wide-range image $\textbf{I}_B$, which is obtained by appropriately correcting, stitching and blending the individual sub-images recovered from the metalens array measurement. To this end, we employ a modified UNet blending network to learn the blending function $f_B$ which takes a blended homography-transformed stack of concatenated $mn$ sub-images (see Sec.~\ref{sec:dataset_synthetic}) and a coarse alpha-blended wide-range image $\textbf{I}_B^\alpha$ as input, and produces the correctly blended image as the output,
\begin{equation}
    \textbf{I}_B = f_B (\{\textbf{I}_{m,n}\}, \textbf{I}_B^\alpha).
\end{equation}
To learn the function $f_B$, the blending network is supervised over groundtruth images acquired using an aberration corrected compound optic camera, see Sec.~\ref{sec:prototype}. The loss function $\mathcal{L}$ used is a combination of pixel-wise error and perceptual loss during the training
\begin{equation}
    \mathcal{L} = \mathcal{L}_1 + \mathcal{L}_\textsc{lpips},
\end{equation}
to allow for accurate reproduction of color and features while also accounting for any misalignments in the in-the-wild captured data pairs as well as systemic errors in the prototype data acquisition setup.
Moreover, supervising our blending network on the full sensor resolution groundtruth image measurements also allows for recovering a high-fidelity latent image from the $m\times n$ low resolution sub-images from the metalens array camera.

\subsection{Implementation}
The proposed deconvolution framework and the learned blending network are implemented in PyTorch. The training for the overall deconvolution approach is done iteratively and progressively to sample over a large plausible manifold of latent images from the sensor measurements. 
For all the training purposes, we use a dataset with groundtruth images of $800 \times 800$ resolution and $9$ patches of $420 \times 420$ sub-images measured from individual metalenses in the nanophotonic array. Training was performed using the paired groundtruth and metalens array measurements acquired using the experimental paired-camera setup, see Sec.~\ref{sec:dataset}. During the deconvolution step, an initial filtered image obtained according to Eq.~\eqref{eq:filter-step} is passed through a probabilistic diffusion-prior model that progressively corrupts the filtered image with additive noise and recovers the latent image by sampling over the manifold of probability distribution of image priors. To preserve color fidelity, we normalize the image to range $[0,1]$. We use the ADAM optimizer with $\beta_1 = 0.5$ and $\beta_2 = 0.999$, and \rev{$\lambda = 1.2$ for the training objective in Eq. ~\eqref{eq:diff_obj}. }

\section{Experimental Prototype and Datasets}
\label{sec:dataset}
This section describes the dataset and camera prototype we use to train the proposed reconstruction network. The training dataset consists of simulated data and captured paired image data. We first describe the synthetic dataset, then the capture setup and the acquisition of the proposed paired dataset. Finally, we describe the  fabrication process of the proposed nano-optical array.

\subsection{Synthetic Dataset}\label{sec:dataset_synthetic}
Training the probabilistic image recovery network described in Sec.~\ref{sec:image_reconstruction} requires a large and diverse set of paired data, which is challenging to acquire in-the-wild. Therefore, we simulate the nanophotonic array camera with the corresponding metalens design parameters, to generate a large synthetic dataset of paired on-sensor and groundtruth measurements. We use this large synthetic dataset for training alongside a smaller real-world dataset for fine-tuning.
Each metalens in the array camera has a focal length of \SI{2}{mm} and covers an FoV of $60^\circ$ for a broadband illumination, with the center-to-center distance between the metalenses on-chip being \SI{2.42}{mm}. Due to the circular aperture of each metaoptic, the sensor measurements exhibit vignetting at higher eccentricities.

For a given groundtruth image, we first crop $9$ images that correspond to the final $3\times 3$ metalens array camera measurement, with each metalens measurement corresponding to $60^\circ$ FoV and the groundtruth image corresponding to a total of $90^\circ$ FoV. Each of the $9$ images are subjected to vignetting where we model the vignetting mask as a fourth-order Butterworth filter with linear intensity fall-off, given by 
\begin{equation}
    \mathbf{V} = {\Bigg( 1+\bigg( \frac{||w||^2}{f_c^2}\bigg)^4 \Bigg)}^{-1}
\end{equation}
where $||\cdot||^2$ denotes the squared magnitude, $w$ is the spatial frequency and $f_c$ is the cutoff frequency of the filter. All parameters are matched to the experimental setting. Note that we apply this filter on each individual metalens measurement only as an intensity mask to the sensor image and our cutoff frequency corresponded to $45^\circ$ of the metalens FoV. The vignetted images are convolved with the simulated PSFs on the sensor as described in Eq.~\eqref{eq:forward_model} and further corrupted by simulated sensor noise described in Eq.~\eqref{eq:sensor-measurement}. 
The simulated individual metalens measurements are then resized and arranged in a $3 \times 3$ array to simulate the nanophotonic sensor capture. To this end, we first compute homographies between the $9$ local image patches as measured by the real nanophotonic array camera and the ground truth compound optic camera, which is described next in Sec~\ref{sec:prototype}, to transform the ground truth image to map that of the sensor capture. We then utilize these homography transforms to project each of the 9 simulated metalens measurements onto the appropriate local patch on the sensor
\begin{equation}
    \mathbf{\hat{p}}^{gt}_{mn} \to \mathbf{H}_{mn}\mathbf{p}^s_{mn} ,
\end{equation}
where $\mathbf{\hat{p}}^{gt}_{mn}$ denotes the coordinates in the ground truth image corresponding to the FoV as captured by the $(m,n)-th$ metalens in the array camera, $\mathbf{p}^s_{mn}$ denotes the sensor coordinate corresponding to the $(m,n)$-th metalens measurement and $\mathbf{H}_{mn}$ denotes the corresponding homography. 
The final sensor measurement is simulated as 
\begin{align}
\mathbf{S}_{mn} & = (\mathbf{H}^{-1}_{mn} \mathbf{I}*\mathbf{V}) \otimes \mathbf{k}_{mn} + \eta_\textsc{sensor},\\
    \mathbf{S} & = \sum_{m,n} \mathbf{S}_{mn} \hspace{2mm} 
    s.t. \hspace{2mm} (m,n)\in \{0,1,2\}; \hspace{2mm} m+n \leq 2 ,
\end{align}
where $\mathbf{S}_{mn}$ denotes the $(m,n)$-th array measurement on the sensor, $\mathbf{S}$ being the final sensor measurement, $\mathbf{H}^{-1}_{mn}$ and $\mathbf{k}_{mn}$ being the corresponding inverse homography and PSF, respectively. 
The sensor noise added is as determined by the parameters $\mathcal{C}_\textsc{sensor} = \{ \sigma_g, a_p\}$ which we determine to be $\sigma_g = 1\times10^{-5}$ and $a_p = 4\times 10^{-5}$ using the calibration method as described in Foi et al.~\cite{Foi2008PracticalPN}.

To generate the full synthetic dataset, we randomly sample 10,000 images from a combination of ImageNet~\cite{deng2009imagenet} and MIT 5K~\cite{fivek} datasets for groundtruth images. Our training dataset contains 8000 images and the validation and test data splits contain 1000 each. 
The networks trained on our synthetic dataset are then further finetuned on in-the-wild real data which we describe in the following. 

\subsection{Experimental Setup and Dataset} 
\label{sec:prototype}
\begin{figure}[t!]
	\centering
		\includegraphics[width=0.48\columnwidth]{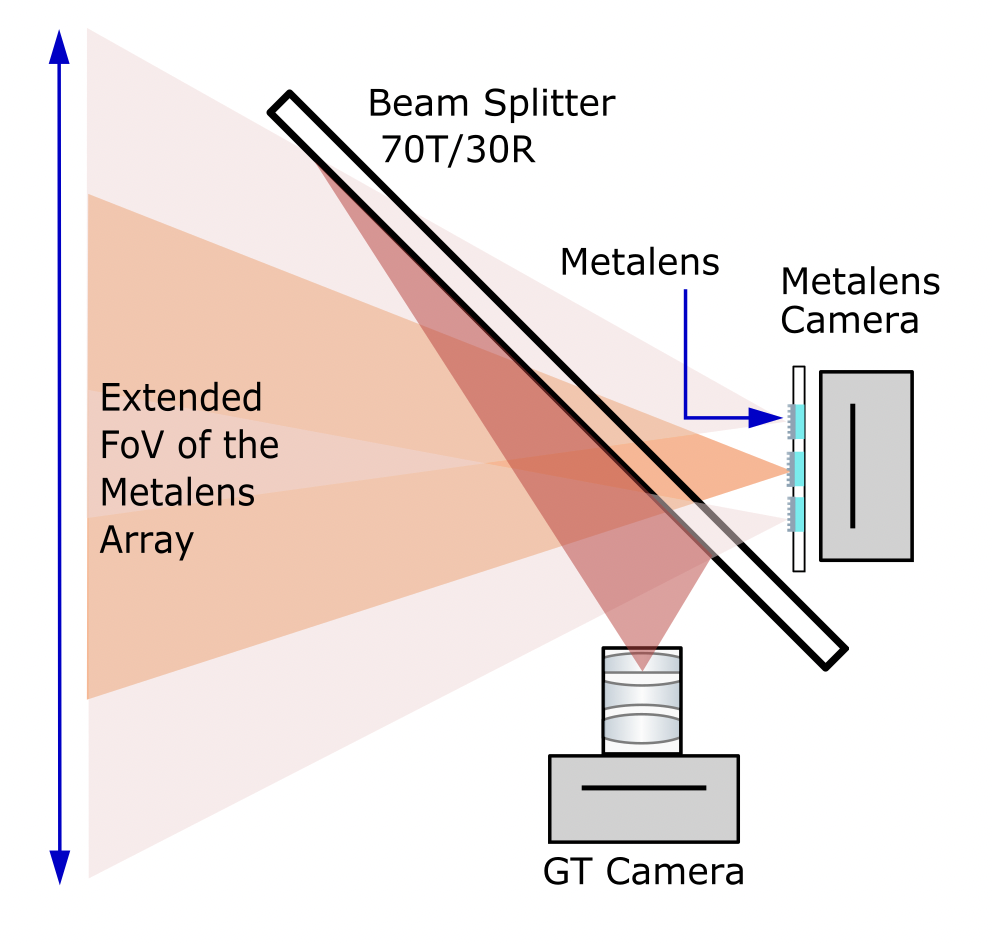}
              \includegraphics[width=.45\columnwidth]{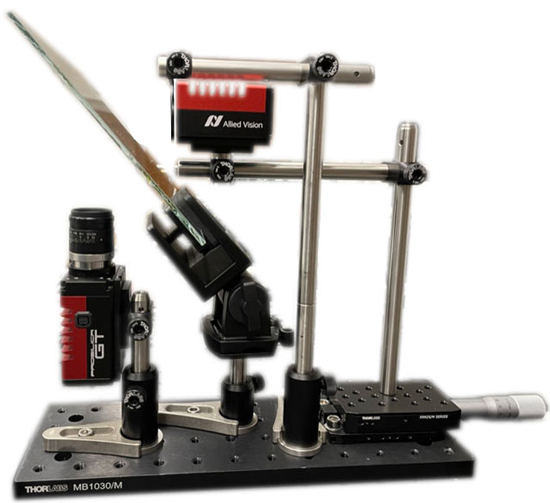}
		\caption{Paired Data Acquisition. In our capture setup, we employ a plate beam splitter, which splits world light into two optical paths by 70\% transmission and 30\% reflection such that the setup can simultaneously capture real-world scenes with one camera in the transmission path that employs the designed metalens array and another camera in the reflection path that employs a conventional off-the-shelf lens (GT camera). The two cameras are aligned and calibrated to map one to the other captures, see text for details.}
		\label{fig:capture_setup}
\end{figure}

To acquire the paired experimental data, we developed a hardware setup shown in Figure~\ref{fig:capture_setup}, which can simultaneously capture real-world scenes from the metalens array camera and a reference camera that has a conventional off-the-shelf lens. To this end, we use a plate beam splitter, which splits world light into two optical paths by 70\% transmission and 30\% reflection. The positions of the plate beam splitter and the two cameras are precisely aligned such that the optical centers and the optical axes of the two cameras are as close as possible. The two cameras are synchronized to capture scenes with the same timestamps.

We employ an Allied Vision GT1930 C sensor of 5.86 micron pixel pitch and 1936x1216 resolution for the metalens array camera such that the effective FoV (Field-of-View) from all the metalens elements in the array can be captured in the same frame. The same sensor is used for the reference camera which has a 3.5~mm focal length, wide FoV lens from Edmund Optics such that we can achieve a FoV larger than the full FoV of the metalens array camera in the ``ground truth'' captures. A third Allied Vision GT1290C camera with 3.75 micron pixel pitch and 1280x960 resolution is used for mounting the metalens proposed by Tseng et al.~\shortcite{neural_nanooptics}, which we compare against in Section~\ref{sec:assessment}. We use Precision Time Protocol (PTP) to synchronize all the cameras such that the captures are taken at the same timestamps with sub-millisecond precision. After we align the sensor parallel to our fabricated metalens array, we perform fine alignment between the sensor and the metalens array with a 3D translation stage, where the sensor is mounted to. When the alignment is completed, the sensor captures the effective FoV of all the metalens array elements and the images are focused on the sensor plane. See Supplemental Material for details.


After the alignment, we conduct PSF measurements of the individual metalens elements in the array, which are used in the model-based part of the image reconstruction method. The light sources that we use are red, green, and blue fiber-coupled LEDs from Thorlabs (M455F3, M530F2,	
and M660FP1). The fiber has a core size of 800 microns diameter and the fiber tip is placed 340 mm away from the metalens array such that it can be approximated as a point source with an angular resolution that is the same as the angular resolution of one pixel in the captured metalens images (\~8 arc-min). The PSFs of all the metalens elements are captured in the same frame. By turning on and off each individual color LED, we can acquire the PSFs of different colors. When alternating between colors, we change the input of the fiber without introducing mechanical shifts to the output of the fiber such that the position of the point light source is fixed.

Next, we align the optical center and the optical axis of the central element from the metalens array camera to those of the reference camera. We use collimated laser and pinhole apertures to make sure the beam splitter is positioned at a 45$^\circ$ tilting angle. Then, we set up the position of the metalens array camera and adjust the laser beam height such that the transmission path is incident on the center metalens element. The center of the reference camera is positioned in the reflection beam path and the distance between the beam splitter and the reference camera sensor is adjusted to the same as that between the beam splitter and the metalens array camera. We achieve accurate alignment by observing a reference target with both cameras simultaneously until the two cameras are aligned.

After all the alignment is completed, the setup is mounted on a tripod with rollers, as shown in Figure~\ref{fig:capture_setup}, such that it can be moved around indoors and outdoors for acquiring a diverse dataset. In the capture process, the exposure time is chosen so that the photos from the two cameras are bright but unsaturated, and the frame rate is chosen to make sure that there is a sufficient difference in the scenes for a few consecutive frames. When we change the scenes, the exposure time of the two cameras is adjusted proportionally such that the image brightness is adapted to different scenes while the light throughput ratio between the two cameras is unchanged. The exposure time of the metalens array camera ranges between 50~ms to 150~ms and the exposure time of the reference camera is 1.2$\times$ that of the metalens camera.

\paragraph{Per-pixel Mapping between Two Cameras} 
To find the per-pixel mapping between the reference camera and the metalens array camera, we have the two cameras capture red, green and blue checkerboard patterns shown on a large LCD screen and then calibrate the distortion coefficients of the two cameras per color channel. After the image acquisition, we perform image rectification for the captures from both cameras. Then, to account for the difference in camera FoV and the difference in viewing perspectives between each metalens array element and the reference camera, we perform homography-based alignment to map the reference camera captures to the captures from all the metalens array elements.

\subsection{Fabrication of the Meta-Optic} 


The optimized meta-optic design described in Sec.~\ref{sec:method} was fabricated in a 700 nm SiN thin film on fused silica substrate. First a SiN thin film was deposited on a fused silica wafer via plasma-enhanced chemical vapor deposition. The meta-optic array was then written on a single chip via electron beam lithography (JEOL-JBX6300FS, 100 kV) using a resist layer (ZEP-520A) and discharging polymer layer (DisCharge H2O). After development, a hard mask of alumina (65 nm) was evaporated, and subsequently lift-off overnight in NMP at 110 $^{\circ}$C. After a brief plasma clean to remove organic residues, we used inductively-coupled reactive ion etching (Oxford Instruments, PlasmaLab100) with a fluorine based etch chemistry to transfer the meta-optic layout from the hard mask into the underneath SiN thin film. Finally, we created apertures for the meta-optics to exclude unmodulated light that passed through non-patterned regions. These apertures were created through optical direct write lithography (Heidelberg-DWL66) and subsequent deposition of a ~150 nm thick gold film. 
\rev{Our array has a total size of $\sim7mm^2$ with elements $1mm$ in diameter and $F\#2.4$. We avoid optical baffles in our prototype and, to ensure no overlap, instead space out the lenslets over the wafer with $\sim15\%$ of the area being used as apertures. However, note that we do not use the peripheral regions of each sublens; hence, we use non-continuous regions of pixels totaling $\sim40\%$ of the full sensor area. In the future, integrating optical baffles to separate array elements may eliminate the need for separation. However, fabricating and aligning baffles is not a simple feat and we prototype our camera without them.}
Please refer to the Supplemental Material for additional details. 

\section{Assessment}
\label{sec:assessment}

\begin{figure*}[t!]
\vspace{-8pt}
    \small
    \renewcommand{\arraystretch}{4.06}
    {\sffamily
  \begin{tabular}{p{2.6cm}p{14.8cm}}
      \specialrule{\heavyrulewidth}{0pt}{0pt}  
      Measurement \& PSF &\multirow{6}{14.6cm}{\includegraphics[width=14.977cm]{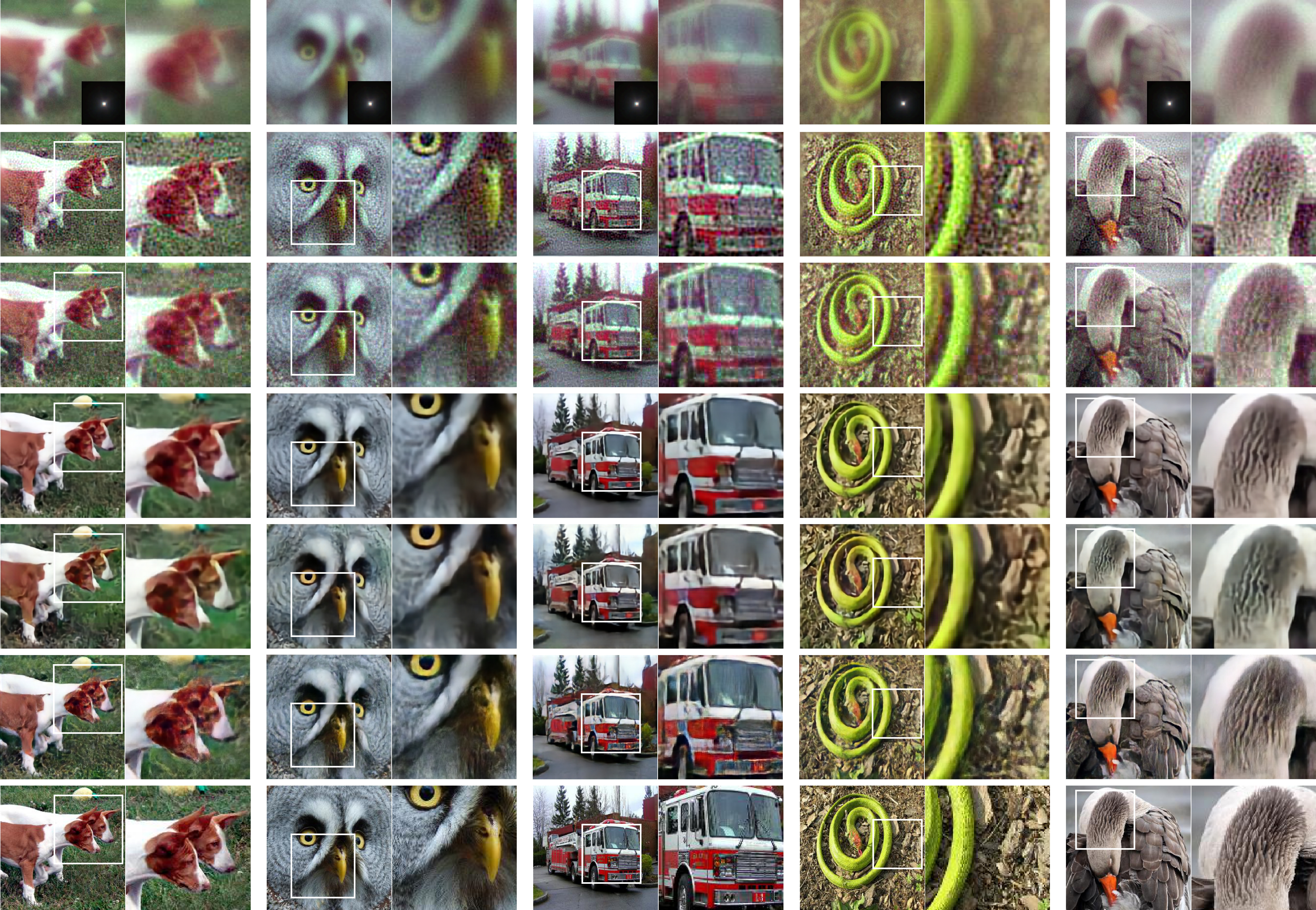}}\\
    Wiener~\shortcite{wiener1949extrapolation}&\\
    Richardson-Lucy \shortcite{richardson1972bayesian,lucy1974iterative}&\\
    \text{FLAT-Net~\shortcite{khan2020flatnet}}&\\
    \text{Multi-Wiener-Net~\shortcite{yanny2022deep}}&\\
    \textbf{Proposed}&\\
    Ground Truth&\\
    \specialrule{\heavyrulewidth}{0pt}{0pt}  
  \end{tabular}
  }
  \caption{Synthetic Qualitative Assessment of Diffusion-based Deconvolution. Results on \emph{unseen} validation set. The two conventional deconvolution approaches (Wiener~\cite{wiener1949extrapolation} and Richardson-Lucy~\cite{richardson1972bayesian}) suffer from apparent reconstruction noise. The predictions from Flatnet~\cite{khan2020flatnet} and Multi-Wiener-Net~\cite{yanny2022deep} are overly smooth with high-frequency details missing. The proposed probabilistic reconstruction method is capable of recovering fine details, such as the grass (left), feathers (center left), or grill (center)  without severe reconstruction noise. }
  \label{fig:results_simulation_single_image}
\end{figure*}

\subsection{Synthetic Evaluation}

Before validating the proposed method on experimental captures, we separately evaluate the probabilistic deconvolution method and the proposed thin camera in simulation. To this end, we use unseen test set (consisting of 1000 images) from our synthetic dataset described in Sec.~\ref{sec:dataset_synthetic} to assess the method with paired ground truth data.

\paragraph{Assessment of Probabilistic Deconvolution.}  
Existing non-blind deconvolution methods do not operate on several sub-aperture images that are combined together to form a final image. To assess the proposed probabilistic deconvolution method in isolation, and allow for a fair comparison, instead of considering all nine sub-apertures of the proposed meta-optic, we consider only the central portion. Doing so allows us to compare the proposed reconstruction method with a single PSF and image -- the setting that existing non-blind deconvolution methods are addressing. For this experiment, we drop the blending operator from the proposed method described in Sec.~\ref{sec:image_reconstruction} and train the remainder of the method as described next. 

\begin{figure*}[ht!]
\vspace{-4pt}
    \small
    {\sffamily
    \setlength{\tabcolsep}{0pt}
    \renewcommand{\arraystretch}{1}
  \begin{tabular}{C{0.2\textwidth}C{0.2\textwidth}C{0.2\textwidth}C{0.2\textwidth}C{0.2\textwidth}}
    Proposed & FlatCam~\shortcite{Asif2017FlatCam} & DiffuserCam~\shortcite{kuo2017diffusercam} & Unrolled Optimization for DiffuserCam~\shortcite{Kingshott:22} & Ground Truth \\
    \multicolumn{5}{c}{\includegraphics[width=\linewidth]{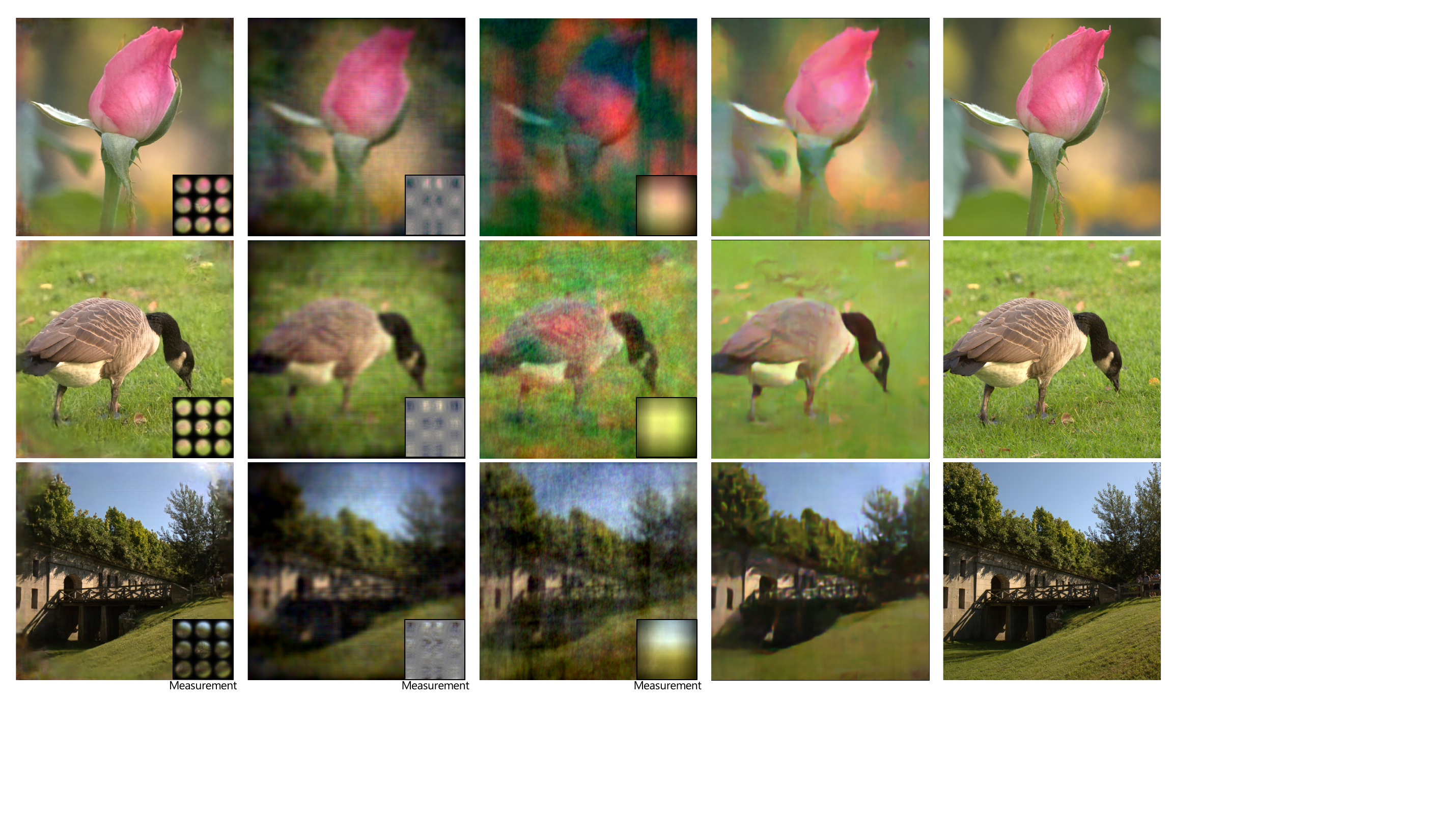}}\\[-12pt]
  \end{tabular}
  }
  \caption{Synthetic Assessment of Thin Cameras. Results on \emph{unseen} validation set. Alternative thin sensing approaches in FlatCam~\shortcite{Asif2017FlatCam} and DiffuserCam~\cite{kuo2017diffusercam} allow for capturing rays from a large cone of angles, however, they mix spatial and color information in PSFs with support of the entire sensor, see insets. This makes the recovery of high-frequency details challenging, even for learning-based methods~\cite{Kingshott:22}. The proposed camera design captures high-quality information across almost the entire field of view. Zoom in to the electronic version of this document for details.}
  \label{tab:results_sim_flat_cameras}
\end{figure*}

We report qualitative and quantitative results in Table~\ref{tab:metric_table} and  Figure~\ref{fig:results_simulation_single_image} 
which both validate the proposed reconstruction method. Specifically, we compare our method to existing conventional and learned non-blind deconvolution methods, that is Wiener inverse filtering~\shortcite{wiener1949extrapolation} and Richardson Lucy iterations~\shortcite{richardson1972bayesian,lucy1974iterative} as traditional methods, and the FLAT-Net~\cite{zhou2020flat} and Multi-Wiener Net~\cite{yanny2022deep} as recent learning-based approaches. We retrain the two learning-based approaches on our synthetic data for a fair comparison. Table~\ref{tab:metric_table} confirms that the proposed method outperforms all compared methods in all metrics, that is SSIM, PSNR, and LPIPS~\cite{zhang2018unreasonable}. Although all learned methods are trained on the same data, the proposed method improves on the existing learned baseline methods by a margin of more than 1~dB in PSNR. The qualitative results reported in Figure~\ref{fig:results_simulation_single_image} confirm this trend. While the learned approaches, FLAT-Net and Multi-Wiener-Net are robust to noise and function in presence of large blur kernels such as the one simulated in this experiment -- in contrast to the conventional methods --  their predictions tend to be oversmoothed. The proposed probabilistic prior is capable of recovering fine details, such as the swan feather structure (right), headlights of the fire truck (center) and details in grass and face of the dogs (left).

\begin{table}[t!]
    \setlength{\tabcolsep}{0em}
    \centering
    \footnotesize
    \begin{tabularx}{\linewidth}{m{0.15\linewidth}XXXXXX}
    \toprule
    &
    {\footnotesize Wiener } &
    {\footnotesize Richardson-Lucy} &
    {\footnotesize Flatnet \footnotemark\shortcite{khan2020flatnet}} &
    {\footnotesize Multi-Wiener-Net \shortcite{yanny2022deep}} &
    {\footnotesize Proposed} \\
    \midrule
    SSIM $\uparrow$ & 
    0.452 & 
    0.486 & 
    0.679 & 
    0.648 & 
    \textbf{0.754} \\
    PSNR [dB] $\uparrow$ & 
    19.38 & 
    19.97 & 
    24.63 & 
    22.90 & 
    \textbf{25.80} \\

    1-LPIPS $\uparrow$ & 
    0.360 & 
    0.495 & 
    0.672 & 
    0.569 & 
    \textbf{0.773} \\
    \bottomrule
\end{tabularx}
    \caption{
        Qualitative Assessment of Probabilistic Non-blind Deconvolution. To evaluate the proposed reconstruction method, we simulate aberrated and noisy images of the central lens in our optical design, see Sec.~\ref{sec:dataset_synthetic}. We evaluate all methods on our synthetic validation set and find that the proposed method outperforms all baselines in SSIM, PSNR, and LPIPS~\shortcite{zhang2018unreasonable}.
        }
    \label{tab:metric_table}
    \vspace{-15pt}
\end{table}

\paragraph{Validation of Thin Imager Design.} 

Next, we validate the proposed thin camera design in simulation. We again rely on the unseen test set from our synthetic dataset described in Sec.~\ref{sec:dataset_synthetic} to evaluate the method with ground truth data available. We now consider all nine sub-apertures on the sensor that require employing the blending operator we dropped for the experiments described above. Figure~\ref{tab:results_sim_flat_cameras} includes simulated sensor measurements as insets for a few scenes, see Supplemental Document for additional simulations.

The qualitative and quantitative evaluations in Table~\ref{tab:sim_cameras_table} and  Figure~\ref{tab:results_sim_flat_cameras}  validate the proposed thin camera design. Here, we compare the proposed thin imager to successful imaging methods with a flat form factor: the FlatCam~\shortcite{Asif2017FlatCam} design, which employs an amplitude mask placed in the sensor cover glass region instead of a compound lens, and DiffuserCam~\shortcite{kuo2017diffusercam} relying on a caustic PSF resulting from a diffuser placed above the coverglass. In addition to evaluating the image formation and reconstruction methods proposed in the original works, we also evaluate recent learning-based reconstruction methods, including FlatNet~\cite{khan2020flatnet} which is capable of learning from FlatCam observations, and the unrolled optimization method with neural network prior from Kingshott et al.~\shortcite{Kingshott:22} that recovers images from diffuser measurements. 
\begin{table}[t!]
    \setlength{\tabcolsep}{0em}
    \centering
    \footnotesize
    \begin{tabularx}{\linewidth}{m{0.15\linewidth}XXXXXX}
    \toprule
    &
    {\footnotesize FlatCam \shortcite{Asif2017FlatCam} } &
    {\footnotesize FlatNet \shortcite{khan2020flatnet} } &
    {\footnotesize DiffuserCam \shortcite{kuo2017diffusercam} } &
    {\footnotesize Kingshott et al. \shortcite{Kingshott:22}} &
    {\footnotesize Proposed~w/ Tikhonov} &
    {\footnotesize Proposed } \\
    \midrule
    SSIM $\uparrow$ & 
    0.544 & 
    0.533 & 
    0.479 & 
    0.594 & 
    0.731 &
    \textbf{0.892} \\
    PSNR [dB] $\uparrow$ & 
    19.25 & 
    20.61 & 
    16.42 & 
    21.24 & 
    25.47 &
    \textbf{32.66} \\

    1-LPIPS $\uparrow$ & 
    0.292 & 
    0.426 & 
    0.255 & 
     0.348 & 
     0.71 &
    \textbf{0.803} \\
    \bottomrule
\end{tabularx}
    \caption{
        \rev{Qualitative Assessment of Thin Camera Design. To evaluate the nanophotonic array camera design proposed in this work, we simulate aberrated and noisy images for our $3 \times 3$ array following Sec.~\ref{sec:dataset_synthetic}, and recover images with the proposed probabilistic reconstruction method. We assess the image quality compared to FlatCam~\cite{Asif2017FlatCam} and DiffuserCam~\cite{antipa2018diffusercam} as alternative thin camera design approaches. We evaluate all methods on our \emph{unseen} synthetic validation set and find that the proposed design compares favorably in SSIM, PSNR, and LPIPS~\shortcite{zhang2018unreasonable}.}
        }
    \label{tab:sim_cameras_table}
    \vspace{-15pt}
\end{table}
We retrain the learning-based approaches on our synthetic data for a fair comparison. The proposed thin imager improves on all alternative designs both quantitatively and qualitatively. While FlatCam and DiffuserCam sensing allow the capture of rays from a large cone of angles, the spatial and color information is entangled in PSFs with support of the entire sensor, making the recovery of high-frequency content challenging independent of the FoV. 
As such, all examples in Figure~\ref{tab:results_sim_flat_cameras} confirm the trend from the quantitative Table~\ref{tab:sim_cameras_table} --- the proposed metasurface array imager is able to image fine details across almost the entire field of view. The flower (top) and draw bridge (bottom) are reconstructed with high fidelity across the entire image. Only in the peripheral corners of the image, the proposed method is not capable of recovering details as image information is missing, which is filled in by the blending network. \footnotetext{We compare here against the generalized version of Flatnet~\shortcite{khan2020flatnet} which does not require separable point-spread functions.}

\rev{The proposed camera design benefits from both the optical design and the probabilistic prior. To analyze the contribution of these two components, we conduct an ablation experiment by replacing the diffusion prior with a non-learned prior. Because spatial priors, including Total Variation (TV) regularization and neural network-based learned priors, both can ``hallucinate'' frequency content missing in the measurements (e.g., high-frequency edges in the case of TV), we compare our approach to Tikhonov regularization \cite{golub1999tikhonov} as a traditional per-pixel prior. We observed an average PSNR of 25.5~dB, which still outperforms all alternative flat camera designs by more than 4~dB. The proposed diffusion prior further improves this by 7.2dB with the same input data used by all methods. These additional evaluations further validate both the optical design and effectiveness of the diffusion prior.
}

\begin{figure}[t]
  \vspace*{-6pt}
	\centering
		\includegraphics[width=1.0\columnwidth]{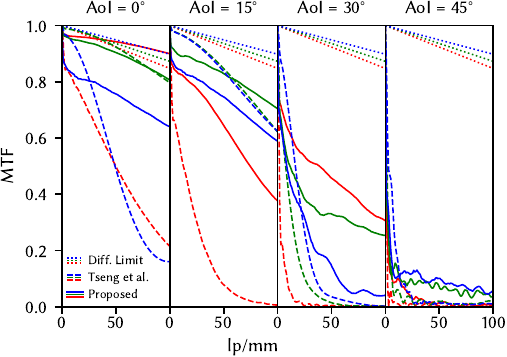}\vspace*{-6pt}
		\caption{MTF for different angles of incidence. The calculated MTF for various angle of incidence (AoI) for array MO with varying wedge phase profile. Thick solid lines correspond to the diffraction limit. Dashed lines correspond to calculated MTF curves for Tseng et al.~\shortcite{neural_nanooptics}. Full thin lines correspond to the MO, presented in this work, with wedge phase as specified in the title. The x-axes represent line pairs per mm, within the specific resolution range for the used sensor, with a pixel size of 5.7 $\mu$m.}
		\label{fig:mtf-analysis}
\end{figure}

\vspace{-1mm}
\subsection{MTF and FoV Analysis}
Next, we analyze the optical performance of the proposed nanophotonic array lens via its theoretical modulation transfer function (MTF), \emph{i.e.}, the ability of the array lens to transfer contrast at a given spatial frequency (resolution) from the object to the imaging sensor. As discussed in Sec.~\ref{sec:method}, our lens is optimized for broadband illumination across the visible spectrum and to span an effective FoV of $70^\circ$ for a $3\times3$ and an FoV of $80^\circ$ for a $5\times 5$ metalens array, respectively, with each individual lens in the array capturing a total FoV of $45^\circ$. We calculate the MTF of our array designs and compare to the the recent design from Tseng et al.~\shortcite{neural_nanooptics}, which is reported to achieve a total FoV of $40^\circ$.

\begin{figure}[t!]
	\centering
              \includegraphics[width=\columnwidth]{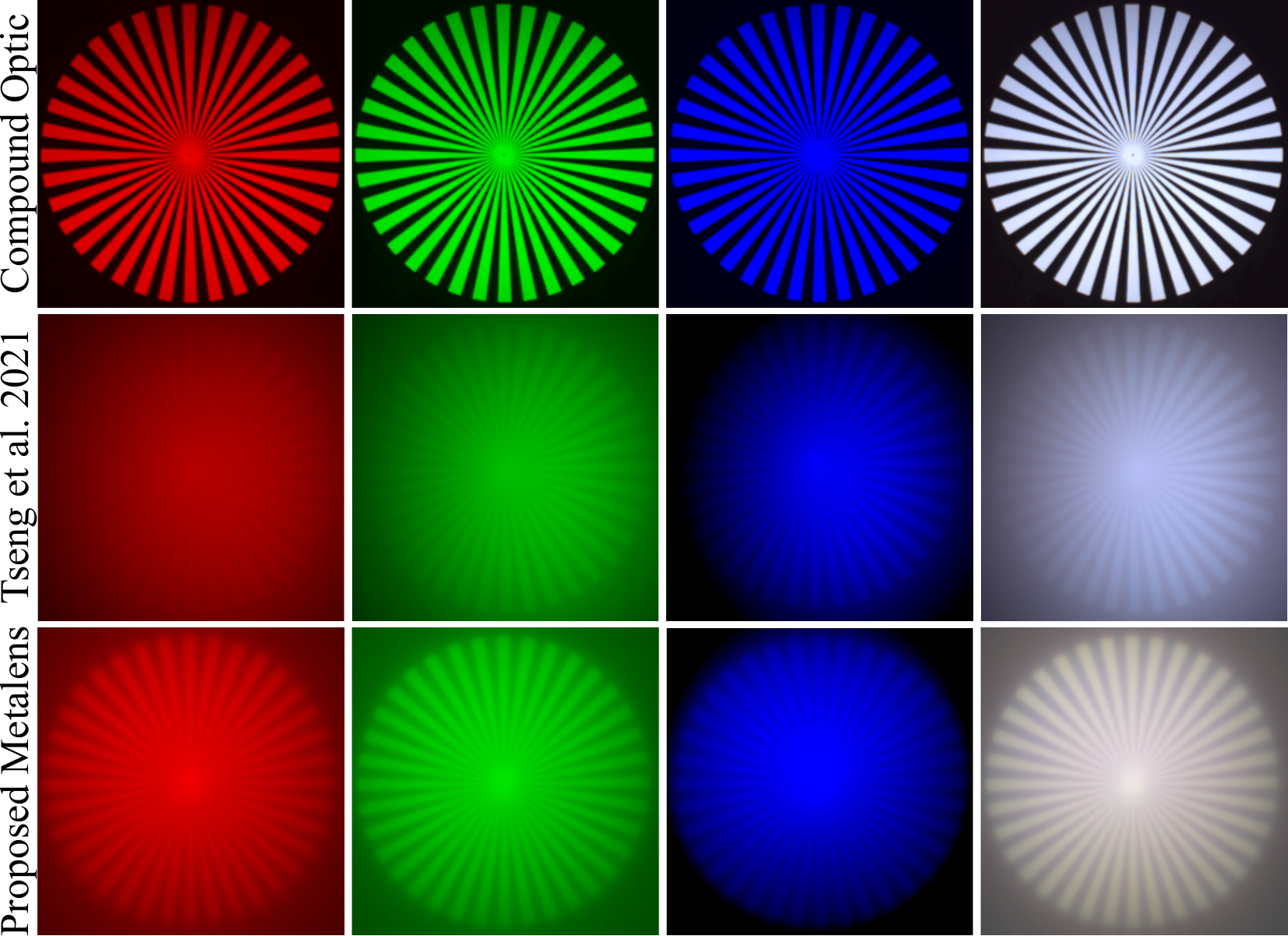}
              \caption{Sensor measurement of the Siemens Star calibration pattern. Even without reconstruction, our metalens design significantly reduces scattering compared to the previous state-of-the-art design proposed by Tseng et al.~\shortcite{neural_nanooptics} across all wavelengths, thereby allowing for broadband imaging in-the-wild. When combined with probablistic deconvolution method, the proposed nanophotonic array camera robustly recovers the latent image.}
		\label{fig:siemens_calibration}
  \vspace{-10pt}
\end{figure}

The analysis in Figure~\ref{fig:mtf-analysis} validates that the proposed metalens exhibits a significantly improved MTF under different angle of incidence (AoI) compared to the existing designs, approaching the efficiency of diffraction limit for normal incidence of light (shown in black in Figure~\ref{fig:mtf-analysis}) and achieving higher MTF values for larger incidence angles. Importantly, even at large AoI, the MTF is sufficiently large to enable reconstruction through a computational backend. This allows for high-fidelity signal measurements and robust full-color deconvolution. 
With increasing angles of incident light, although the MTF performance drops, our metalens compares favorably to the design from Tseng et al. The MTF performance of our lens is also reflected in the raw sensor measurements. Figure~\ref{fig:siemens_calibration} reports the raw sensor measurement using a compound optic, the metalens by Tseng et al.~\shortcite{neural_nanooptics}, and the measurement from our proposed metalens design. Although Tseng et al.~\shortcite{neural_nanooptics} designed their metalens specificially for trichromatic red, green and blue wavelengths, the sensor measurement of the calibration Siemens star pattern show significant scattering resulting in loss of image quality. Note that the MTF considerations above are based on the direct sensor measurement without the algorithmic framework employed for latent image recovery. However, the effective MTF which also considers the image recovery algorithm~\cite{fontbonne2022end} which we review later in our experimental results. 



\subsection{Experimental Assessment}

\begin{figure*}[t!]
	\centering

\small
    {\sffamily
    \setlength{\tabcolsep}{0pt}
    \renewcommand{\arraystretch}{1}
  \begin{tabular}{C{0.2\textwidth}C{0.2\textwidth}C{0.2\textwidth}C{0.2\textwidth}C{0.2\textwidth}}
    Array Measurement & 
    Proposed Reconstruction & 
    Tseng et al.~\shortcite{neural_nanooptics} Measurement & Tseng et al.~\shortcite{neural_nanooptics} Reconstruction & Compound Lens Capture Ground Truth \\
    \multicolumn{5}{c}{\includegraphics[width=0.98\linewidth]{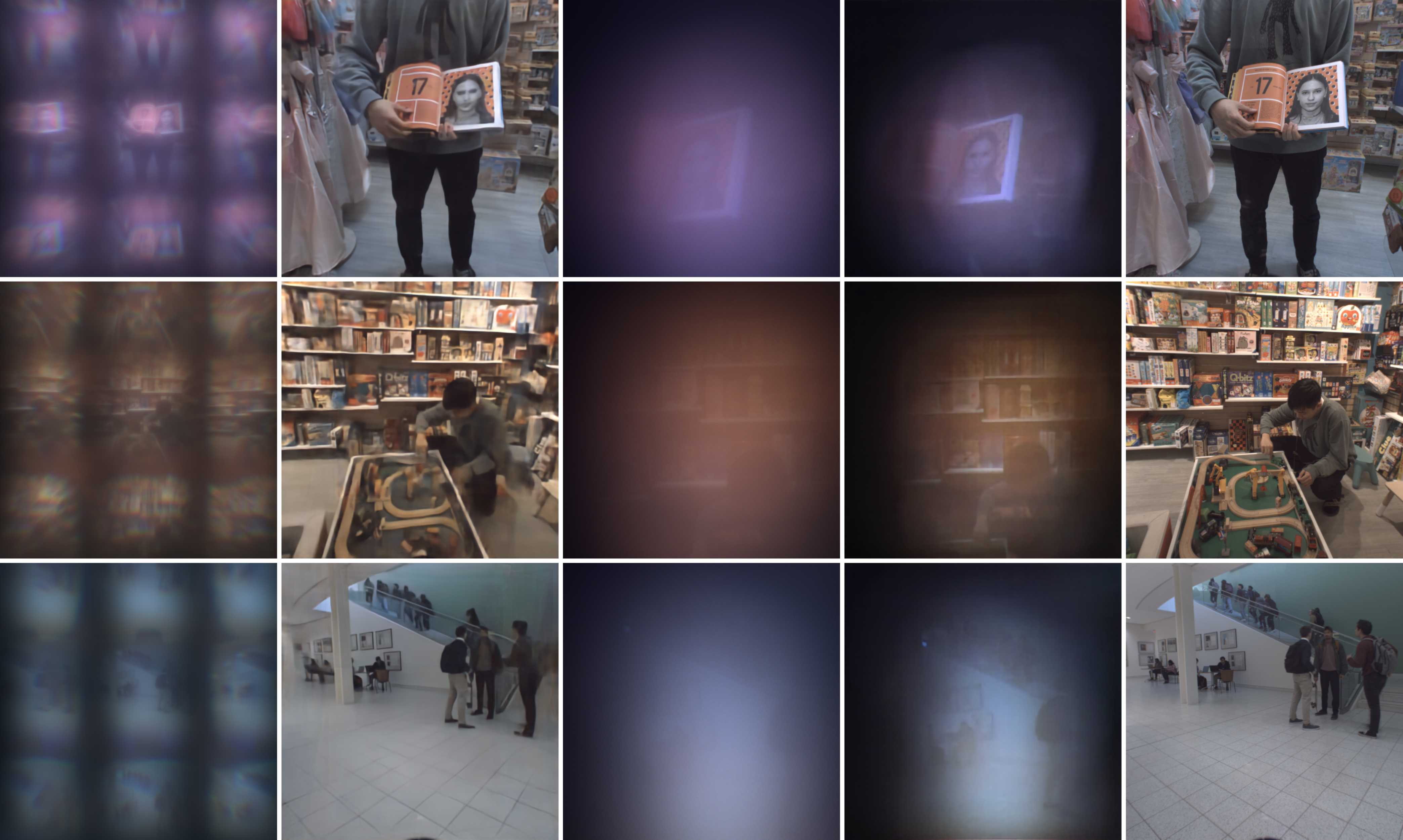}}\\[-12pt]
  \end{tabular}
  }
              
              \caption{Experimental evaluation of the proposed thin nanophotonic camera on broadband indoor scenes. The proposed nanophotonic array optic with the probabilistic deconvolution method reconstructs the underlying latent image robustly in broadband lit environments. As a comparison, Tseng et al.~\shortcite{neural_nanooptics} cannot recover the color and spatial details of the scenes well. See Supplemental Material for lab captures acquired of a narrow-band screen where the proposed design and Tseng et. al both function adequately. }
		\label{fig:experimental_results}
  \vspace{-10pt}
\end{figure*}

\begin{figure*}[t!]
	\centering

\small
    {\sffamily
    \setlength{\tabcolsep}{0pt}
    \renewcommand{\arraystretch}{1}
  \begin{tabular}{C{0.2\textwidth}C{0.2\textwidth}C{0.2\textwidth}C{0.2\textwidth}C{0.2\textwidth}}
    Array Measurement & 
    Proposed Reconstruction & 
    Tseng et al.~\shortcite{neural_nanooptics} Measurement & Tseng et al.~\shortcite{neural_nanooptics} Reconstruction & Compound Lens Capture Ground Truth \\
    \multicolumn{5}{c}{\includegraphics[width=0.98\linewidth]{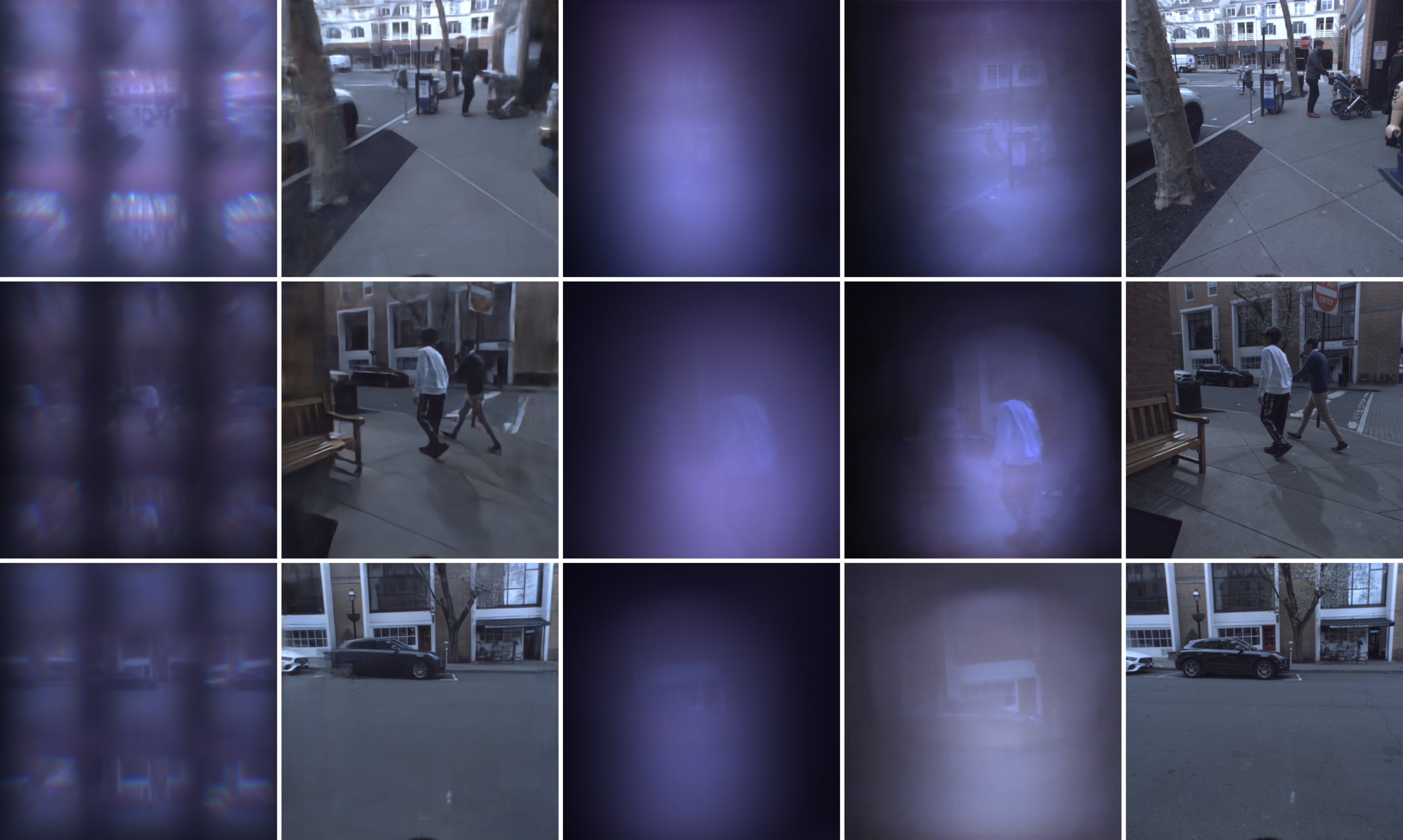}}\\[-12pt]
  \end{tabular}
  }
              
              \caption{Experimental evaluation of the proposed thin nanophotonic camera on broadband outdoor scenes. The proposed nanophotonic array optic with the probabilistic deconvolution method reconstructs the underlying latent image robustly in broadband lit environments, outperforming Tseng et al.~\shortcite{neural_nanooptics} significantly in outdoor scenes. See Supplemental Material for lab captures acquired of a narrow-band screen where the proposed design and Tseng et. al both function adequately.}
		\label{fig:experimental_results_2}
  \vspace{-10pt}
\end{figure*}

In the following, we validate the proposed camera design with experimental reconstructions from captures acquired with the prototype system from Sec.~\ref{sec:prototype}. To this end, we aim to capture scenes that feature high-contrast detail, depth discontinuities, and color spanning a large gamut. To test the camera system in in-the-wild environments, we acquire scenes in typical broadband indoor and outdoor scenarios. As such, we note that, to the best of our knowledge,  ours is the \emph{first demonstration of broadband nanophotonic imaging outside the lab}. Figure~\ref{fig:experimental_results} and Figure~\ref{fig:experimental_results_2} shows images as captured by the previous state-of-the-art~\cite{neural_nanooptics} and proposed thin-lens camera, and the corresponding reference images captured using a compound optical lens for a variety of indoor and outdoor scenes. The reconstructions from Tseng et al. were measured on a sensor of smaller size, see Sec~\ref{sec:prototype}, which we do not resize here, unlike in the Teaser~\ref{fig:teaser}.3 The images recovered using the proposed nano-optic and reconstruction algorithm outperform existing approaches. The proposed thin camera is capable of imaging the scene adequately with accurate color reproduction. While the peripheral regions, reconstructed from elements with strong wedge phase elements, contain less spatial detail than the center region, they still recover detail over the entire design field of view. The center region of the recovered images has relatively high image quality and captures fine detail present in the reference images. The reconstructed images suffer from no apparent chromatic aberrations, which have been an open problem in the design of broadband metasurface optics.

\paragraph{Comparison to Neural Nano-Optics~\cite{neural_nanooptics}}

We compare the proposed design experimentally to the broadband design from Tseng et al.~\shortcite{neural_nanooptics}. While their lens design is the most successful existing broadband metalens design, it is designed for a fixed set of three wavelength bands. As Tseng et al.~\cite{neural_nanooptics} report, their design performs well for the narrow selective spectrum of an OLED display that is imaged with an optical relay system. We confirm this experimentally in the Supplemental Material. For the full broadband scenarios that we tackle in our work, their design comes with severe scattering that is not apparent when imaging a screen with black surrounding region, as shown in Figure~\ref{fig:experimental_results} and Figure~\ref{fig:experimental_results_2}. For the experimental scenes captured in our work, the proposed method significantly outperforms this existing design in image quality and size of the field of view. This experiment validates the proposed broadband design methodology and the array sensing approach we investigate in this work.

\paragraph{Experimental Validation of Denser 5~$\times$~5 Design}

\begin{figure}[]
	\centering
              \includegraphics[width=\columnwidth]{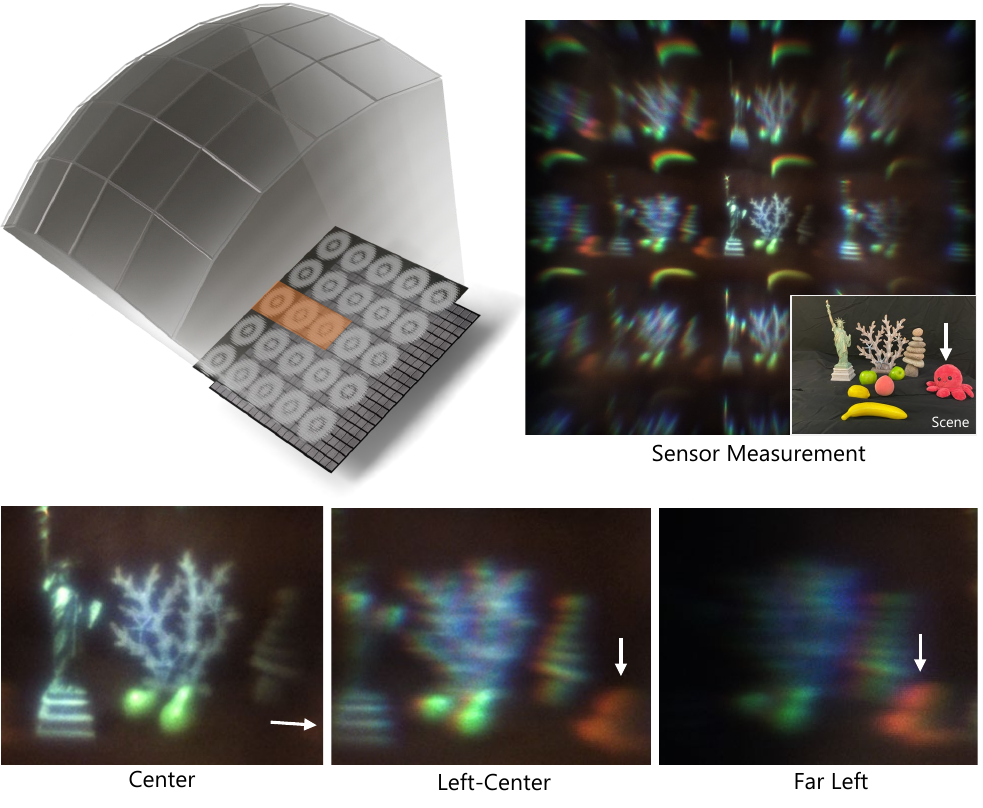}
		\caption{Experimental Assessment of 5~$\times$~5 Prototype Lens. Although the fabricated array does slightly exceed the available sensor area, see text, the capture of the horizontal set of elements (indicated in orange) illustrates that neighboring elements with increasing wedge angles capture successively oblique field angles. Moving towards the array periphery, the red plush toy (indicated with white arrow) is entering (center-left) and moving progressively towards the center of the field of view of each individual lens.}
		\label{fig:5x5_results}
  \vspace{-10pt}
\end{figure}

In addition to the 3~$\times$~3 array investigated above, we have also fabricated an additional 5~$\times$~5 array with an additional peripheral set of nanophotonic lenslets to cover a larger field of view of 120$^\circ$. Unfortunately, the sensors available to us (with sufficient lead time) were slightly too small to capture the entire array, and spacing the elements closer would have required baffles and the removal of the coverglass on the sensor~\footnote{The epoxy-glued cover glasses on commodity mass-market sensor packages cannot be removed without specialized tools or destroying the sensor.}. Figure~\ref{fig:5x5_results} reports the measurement of a still life scene captured under broadband quartz light \emph{without} image reconstruction. Although the full capture of the array is cropped due to the available sensor size, we can observe the increasing field of view from the center element to the periphery. Validating the proposed design, as indicated in the figure, the red plush toy, completely outside the field of view of the center element, enters the field of view of the middle, and moves to the center in the left element.

\section{Conclusion}
\label{sec:conclusion}

We investigate a flat camera that employs a novel array of nanophotonic optics that are optimized for broadband spectrum and collaboratively capture a larger field of view than a single element. The proposed nanophotonic array is embedded on a metasurface that sits on top of the sensor cover glass making the proposed imager thin and manufacturable with a single-element optical system. Although we devise a differentiable lens design method for the proposed array metasurface sensor -- allowing us to suppress aberrations across the full visible spectrum that exist in today's heuristic and optimized metasurface optics -- the proposed design is not without aberrations. We propose a probabilistic image reconstruction method that allows us to recover images in presence of scene-dependent aberrations in broadband -- an open problem using metasurface optics. We validate the proposed nanophotonic array camera design experimentally and in simulation, confirming the effectiveness not only of the optical design, compared against existing broadband metasurface optics, but also the deconvolution method, compared in isolation or against alternative thin camera designs. 
\rev{
In the future, we plan to explore integrating low-cost baffles and the co-design with sensor color-filter arrays into the proposed design which requires scale-able fabrication integrated into the sensor cover glass.
We hope that the proposed camera can not only inspire novel designs, e.g., flexible sensor arrays, but also re-open an exciting design space computational photography community has explored in the past, that is light field arrays, color-multiplexed arrays,and task-specific array optics -- all now directly on the sensor.}



\bibliographystyle{ACM-Reference-Format}
\bibliography{references}
\end{document}